\documentclass[12pt]{article}

\usepackage{amsfonts}
\usepackage{bbm}
\usepackage{bm}

\newcommand{\bmat}{\left(\begin{array}}
\newcommand{\emat}{\end{array}\right)}

\def\gtrsim{\mathrel{\raise.3ex\hbox{$>$\kern-.75em\lower1ex\hbox{$\sim$}}}}
\def\NPB#1#2#3{Nucl. Phys. B{#1} (#2) #3}

\def\ts{\textstyle}
\def\p{\partial}
\def\a{\alpha}

\def\b{\beta}
\def\g{\gamma}
\def\d{\delta}
\def\ep{\epsilon}

\def\om{\omega}
\def\r{\rho}

\def\-{\hphantom{-}}
\def\ov{\overline}
\def\s2{\frac{1}{\sqrt2}}

\def\wt{\widetilde}
\def\oh{\frac{1}{2}}
\def\beq{\begin{equation}}
\def\eeq{\end{equation}}
\def\beqa{\begin{eqnarray}}
\def\eeqa{\end{eqnarray}}
\def\D{{\rm D}}

\def\im{{\rm Im \,}}
\def\re{{\rm Re \,}}

\def\T{{\rm T}}
\def\Z{{\mathbb Z}}

\def\cg{{\cal G}}

\def\cw{{\cal W}} 
\def\cq{{\cal Q}}
\def\cj{{\cal J}}

\def\cp{{\cal P}}

\def\cn{{\mathcal N}}

\def\ch{{\cal H}}
\def\cf{{\cal F}}

\def\cx{{\cal X}}

\def\ba{{\bar a}}
\def\bb{{\bar b}}
\def\bh{{\bar h}}
\def\bg{{\bar g}}
\def\bbp{{\bar b}^\prime}

\def\bgp{{\bar g}^\prime}

\def\ff{{\bar f}}
\def\bbet{{\bar \beta}}
\def\bgam{{\bar \gamma}}
\def\mg{m_{3/2}}
\def\mg2{m^2_{3/2}}

\def\deq#1{\mbox{$D$=#1}}
\def\neq#1{\mbox{$\cn$=#1}}
\def\Dsl{\,\raise.15ex\hbox{/}\mkern-13.5mu D} 

\newcommand{\mathsmaller}[1]{\mbox{\footnotesize$#1$}}
\def\bom{{\bm\omega}}
\def\bF{{\bm{F}}}
\def\bR{{\bm{R}}}
\def\aneq{\not=}
\begin{document}
\pagestyle{plain}

\makeatletter
\@addtoreset{equation}{section}
\makeatother
\renewcommand{\theequation}{\thesection.\arabic{equation}}
\pagestyle{empty}
\rightline{CAB-IB/2900206, IFT-UAM/CSIC-06-05}
\begin{center}
\LARGE{More Dual Fluxes and Moduli Fixing 
\\[10mm]}
\large{ G. Aldazabal${}^a$, P.G. C\'amara${}^b$, 
A. Font${}^b$\footnote{On leave from Departamento de F\'{\i}sica, Facultad de Ciencias,
Universidad Central de Venezuela, A.P. 20513, Caracas 1020-A, Venezuela.}
and L.E. Ib\'a\~nez${}^b$ \\[6mm]}
\small{
${}^a$Instituto Balseiro, CNEA, Centro At\'omico Bariloche, \\[-0.3em]
8400 S.C. de Bariloche, and CONICET, Argentina. \\ 
${}^b$Departamento de F\'{\i}sica Te\'orica C-XI
and Instituto de F\'{\i}sica Te\'orica  C-XVI,\\[-0.3em]
Universidad Aut\'onoma de Madrid,
Cantoblanco, 28049 Madrid, Spain 
\\[1cm]} 
\small{\bf Abstract} \\[0.5cm]
\end{center}
{\small
We generalize the recent proposal that invariance
under T-duality leads to additional non-geometric fluxes 
required so that superpotentials in type IIA and type IIB
orientifolds match. 
We show that 
invariance under type IIB  S-duality requires the introduction 
of a new set of fluxes leading to further superpotential terms.
We find new classes of  \neq1 supersymmetric Minkowski vacua based on  
type IIB  toroidal orientifolds in which not only dilaton 
and complex moduli but also K\"ahler moduli are fixed.
The chains of dualities 
relating type II orientifolds to heterotic and M-theory 
compactifications  
suggests the existence of yet further flux degrees of freedom. 
Restricting to a
particular type IIA/IIB or heterotic compactification only
some of these degrees of freedom have a simple perturbative and/or
geometric interpretation.
}


\newpage
\setcounter{page}{1}
\pagestyle{plain}
\renewcommand{\thefootnote}{\arabic{footnote}}
\setcounter{footnote}{0}

\section{Introduction}
\label{seci}

Fluxes of antisymmetric fields in string compactifications 
have been studied intensively in the last few years \cite{grana}. 
One of the most interesting aspects of the presence of these fluxes
is that they may generate superpotential couplings for the 
compactification moduli \cite{gvw}. These superpotentials, perhaps 
supplemented by  other dynamical effects like gaugino condensation, 
may lead to a full determination of all moduli, solving an outstanding 
problem in string theory. 

Type IIA orientifolds with fluxes were rather neglected in
the past but are are starting to receive more attention 
\cite{kk,gl,dkpz,vz,DeWolfe,cfi, Camara, Chen}.
In IIA compactifications it is possible to switch on backgrounds of
even RR and odd NS forms. This in turn implies the important
result that IIA flux-induced superpotentials depend on all geometrical
moduli as well as on the dilaton. 
In particular, it has been shown that in simple toroidal orientifolds one can
stabilize all closed string moduli in AdS space without considering
extra non-perturbative (e.g. gaugino condensation) effects
\cite{dkpz,DeWolfe,cfi}.
Moreover, in type IIA it is natural to incorporate metric fluxes \cite{dkpz, vz}
that correspond to generalized Scherk-Schwarz reductions \cite{ss, km,
DallAgata, Andrianopoli, Reid}. In presence of metric backgrounds, the 
flux contribution to RR tadpoles can have either sign or even
vanish, opening interesting possibilities for model-building \cite{cfi}.   

A logical question is whether type IIB flux-induced superpotentials
can also depend on all moduli.
In fact, it has been recently shown \cite{stw} that in order to
recover T-duality invariance between the type IIA and type IIB
versions of the same compactification in the presence of 
RR, NS and metric backgrounds, new `non-geometric' fluxes
have to be introduced. Once this is done the superpotentials
on the type IIA and type IIB sides adequately match.
Such non-geometric fluxes had already been, and continue to
be, studied by several authors 
\cite{hmw, dh1, kstt, Ozer, fww, Schulz, Hull, Reid, fw, dh2, lsw}.
In this paper we generalize the work of \cite{stw} to orientifolds
with several diagonal geometrical moduli. We will give explicit
expressions for the superpotential and tadpoles in terms of
integrals involving the flux tensors.

As pointed out in \cite{stw}, there is still a puzzle. 
We know that the type IIB theory has
an S-duality symmetry built in. This symmetry is inherited by
the effective potential of type IIB orientifold compactifications
in the presence of standard RR and NS  backgrounds. However the  
symmetry disappears if we introduce the new non-geometric fluxes.
In the present paper we argue that in order to recover
S-duality in the underlying orientifold theory one has
to introduce an extra set of `S-dual' fluxes. These S-dual fluxes give
rise to  new terms in the effective superpotential 
and do also contribute to RR tadpoles and Bianchi conditions.
In order to study
the structure of tadpoles and Bianchi identities
we make   use of $SL(2,\Z)_S$ transformations.
Whereas the `3-brane' RR tadpole is S-duality invariant,
`7-brane' tadpoles come in a $SL(2,\Z)_S$ triplet and 
couple to a triplet of 8-forms in agreement with
results in \cite{mo, dlt, el, brkor}. On the other hand, S-duality
transformations
on Bianchi identities give rise to new constraints.
The extra flux degrees of freedom still respect T-duality
among type IIA and type IIB so that the effective action is
both S-duality and T-duality invariant. 
We describe all these fluxes in the context of a simple 
$\T^6/(\Omega (-1)^{F_L} I_6)$  type IIB orientifold
($\T^6/(\Omega (-1)^{F_L} I_3)$ in type IIA)
and concentrate on the dynamics of the seven diagonal moduli.
The additional S-dual terms in the superpotential allow us to
obtain new classes of  \neq1 Minkowski vacua 
in which not only dilaton
and complex moduli but also K\"ahler moduli are fixed.
As it happened in the AdS type IIA vacua in ref.\cite{cfi},
the contribution of fluxes to RR tadpoles may have
the same or opposite sign to that of D-branes.

We also argue that comparison with related compactifications 
leading to similar  models
(based on M-theory compactified on $X_7=\T^7/\Z_2\times \Z_2\times \Z_2$
and  heterotic on twisted tori) points at the existence of further
flux degrees of freedom beyond those mentioned above. In 
particular,  the combination of type I-heterotic duality with 
heterotic self T-duality suggests that the full
underlying duality symmetry in these toroidal examples 
includes $SL(2,\Z)^7=SL(2,\Z)_S\times SL(2,\Z)_U^3\times SL(2,\Z)_T^3$.
Full duality invariance requires the presence of up to $2^7$
fluxes.  

The structure of this paper is as follows.
In the next chapter we define the geometry of our
orientifold examples and describe the fluxes, superpotential,
RR tadpoles and Bianchi conditions for both 
the type IIA and its T-dual type IIB version.
In chapter 3 we generalize the setting and describe in detail
how T-duality requires the introduction of non-geometric
fluxes and give explicit formulae for the superpotential
in this generalized case. The modifications of
RR tadpole conditions as well as Bianchi identities are
discussed for the three T-dual settings of type IIB with O3-planes,
type IIA with O6-planes, and type IIB with O9-planes.
We also briefly describe some general properties 
of minima of the scalar potential induced by fluxes.
In chapter 4 we describe how the S-duality underlying 
type IIB theory requires the introduction of novel
`S-dual' flux degrees of freedom giving  rise 
to new terms in the superpotential. In order to study
the structure of tadpoles and Bianchi identities 
we make   use of $SL(2,\Z)_S$ transformations and
provide particular solutions of the constraints.
We report on chapter 5 on a search for \neq1 
Minkowski minima of the flux-induced scalar potential
 showing  several examples. 
In chapter 6 we compare the type II results with 
those coming from compactifying M-theory and heterotic strings 
on twisted tori.
We also discuss the generalization to $SL(2,\Z)^7$ 
invariant superpotentials in which fluxes fill an spinorial
representation as described in the appendix. 
Some final comments are left for chapter 7.

\section{Orientifolds with NS, RR and geometric fluxes}
\label{sec1}

Before addressing the issue of non-geometric fluxes as well as
the new fluxes implied by S-duality,
 we review in this chapter the basic features of 
the toroidal orientifolds
under consideration. We start by describing  the moduli and fluxes
of a type IIA  orientifold on $\T^6/[\Omega_P (-1)^{F_L} \sigma_A]$.
In this case the orientifold symmetry allows to include metric fluxes.
We then move to the T-dual  
IIB orientifold on $\T^6/[\Omega_P (-1)^{F_L} \sigma_B]$. For both cases
we give the expressions for the flux-induced moduli superpotential
as well as Bianchi and RR tadpole cancellation conditions.

\subsection{Notation}
\label{ssecnot}

Let us first fix our notation for the geometric moduli on the tori.
We focus on compactifications on a factorized torus 
 $\T^6=\otimes_{i=1}^3 \T_i^2$.
As basis of closed 3-forms with one leg on each sub-torus we take
\beqa
\a_0 & = & dx^1 \wedge dx^2 \wedge dx^3 \quad ; \quad
\b_0 = dy^1 \wedge dy^2 \wedge dy^3 \ , \nonumber \\[0.2cm]
\a_1 & = & dx^1 \wedge dy^2 \wedge dy^3 \quad ; \quad
\b_1 = dy^1 \wedge dx^2 \wedge dx^3 \ , \label{abbasis} \\[0.2cm]
\a_2 & = & dy^1 \wedge dx^2 \wedge dy^3 \quad ; \quad
\b_2 = dx^1 \wedge dy^2 \wedge dx^3 \ , \nonumber \\[0.2cm]
\a_3 & = & dy^1 \wedge dy^2 \wedge dx^3 \quad ; \quad
\b_3 = dx^1 \wedge dx^2 \wedge dy^3 \ , \nonumber
\eeqa
where  $y^i=x^{i+3}$.
Our normalization is $\int_{\T^6} \a_I \wedge \b_J = \d_{IJ}$.
The closed 2-forms and their dual 4-forms are
\beq
\om_i = -dx^i \wedge dy^i \quad ; \quad
\wt{\om}_i = dx^j \wedge dy^j \wedge dx^k \wedge dy^k
\quad ; \ i\not=j\not=k \ .
\label{omomt}
\eeq
Notice that $\int_{\T^6} \om_i \wedge \wt{\om}_j = \d_{ij}$.
Each sub-torus $\T_j^2$ has area $(2\pi)^2 A_j$ and 
the geometric complex structure
parameters are  given by
\beq
\tau_j  =  \frac1{e^2_{jx}}(A_j + i\,  e_{jx} \cdot e_{jy}) 
\label{gmod}
\eeq
where $e_{jx}$ and $e_{jy}$ are the lattice vectors of
sizes $R_x^j$ and $R_y^j$.
The K\"ahler form is
\beq
J=\sum_{i=1}^3 A_i \omega_i  \ .
\label{kform}
\eeq
As usual the holomorphic 3-form can be written as
\beq
\Omega=(dx^1 + i\tau_1\, dy^1) \wedge (dx^2 + i\tau_2\, dy^2)
\wedge (dx^3 + i\tau_3\, dy^3) \ .
\label{holo}
\eeq
Clearly, $\Omega$ can be expanded in the basis of 3-forms.

\subsection{IIA orientifold with O6-planes}
\label{sseciia}

Consider the IIA orientifold on $\T^6/[\Omega_P (-1)^{F_L} \sigma_A]$,
where $\Omega_P$ is the world-sheet parity operator and $(-1)^{F_L}$ is
the space-time fermionic number for left-movers. The
involution $\sigma_A$ acts on the K\"ahler
form and the holomorphic 3-form as $\sigma_A(J)=-J$ and
$\sigma_A(\Omega)=\Omega^*$. In terms of the coordinates  
this is $\sigma_A(x^i)=x^i$ and $\sigma_A(y^i)=-y^i$.
These imply  O6-planes that span space-time and the
$x^i$ directions. Each sub-torus has now a 
square lattice, consistent with the involution. Thus,
$\tau_j=R_y^j/R_x^j$ and $A_j = R_x^j R_y^j$.

We concentrate our analysis on the seven 
diagonal  moduli of this IIA orientifold, the dilaton
$S$, three K\"ahler moduli $T_i$ and three complex structure
moduli $U_i$. As shown in \cite{gl} they can be concisely
described in terms of the complexified forms
\beqa
J_c & = & B + iJ = i
\sum_{i=1}^3 T_i \omega_i \ ,
\nonumber \\[0.2cm]
\Omega_c & = &  C_3 + i\re (C\Omega) = i S \a_0 - i\sum_{i=1}^3 U_i \a_i \  .
\label{iiamoduli}
\eeqa
Here $B$ is the NS 2-form whereas $C_3$ is the RR 3-form that is even
under $\sigma_A$ and can therefore be expanded in the $\a_J$. The
compensator field $C$ is specified by
\beq 
C= e^{-\phi_4}
e^{K_{cs}/2} \quad ; \quad K_{cs}=-\log [-\frac{i}8 \int_{\T^6} \Omega
\wedge \Omega^* ] \ ,
\label{cfield}
\eeq 
where $\phi_4$ is the T-duality invariant four-dimensional
dilaton given by $e^{\phi_4}= e^{\phi}/\sqrt{{\rm vol}\, \T^6}$.
Clearly, $\re T_j = R_x^j R_y^j$ and from $\Omega_c$ we readily find
\beq
\re S = e^{-\phi}  R_y^1 R_y^2 R_y^3  \quad ;
\quad \re U_i = e^{-\phi} R_x^i R_y^j R_y^k 
\quad ; \quad \ i\not=j\not=k \ .
\label{csmoduli}
\eeq
We are measuring all lengths in units of $\sqrt{\a^\prime}$.

The K\"ahler potential for the moduli takes the usual form 
\beq 
K= - \log(S+S^*) - \sum_{i=1}^3 \log(U_i + U_i^*) -
\sum_{i=1}^3 \log(T_i + T_i^*) \ .
\label{kpot}
\eeq
A superpotential is generated by turning on fluxes as we discuss next.

Under the orientifold involution the NS $H_3$ is odd,
the RR forms $F_0$ and $F_4$ are even while $F_2$ and $F_6$ are odd.
Thus the general fluxes allowed are
\beqa 
\ov{H}_3 & = & \sum_{L=0}^3 h_L \b_L  
\ ; \label{hfluxa} \\[0.2cm]
\ov{F}_0 & = & -m  \quad ; \quad
\ov{F}_2 = \sum_{i=1}^3 q_i \om_i \quad  ; \quad
\ov{F}_4 = \sum_{i=1}^4 e_i \wt{\om}_i \quad ; \quad 
\ov{F}_6  =  e_0 \a_0 \wedge \b_0  \ .
\label{ffluxa} 
\eeqa 
The coefficients in these expansions are integers 
since the integrals of the fluxes over the corresponding $p$-cycles are quantized.
To avoid subtleties with exotic
orientifold planes we take the flux integers to be even.
As in \cite{cfi} we take all forms to have
dimensions $({\rm length})^{-1}$ so that moduli
fields are all dimensionless.

The orientifold involution also allows for metric fluxes that are
deformations of the original manifold. Such backgrounds appear
naturally in the context of Scherk-Schwarz reductions \cite{ss}. They are
equivalent \cite{km, DallAgata, Andrianopoli, Reid} to compactification 
on a twisted torus defined by
\beq
d\eta^P = -\oh \omega^P_{MN} \eta^M \wedge \eta^N \quad ;  \quad
M,N,P= 1,\cdots, 6 \  ,
\label{ocon}
\eeq
where $\eta^P$ are the tangent 1-forms. 
The metric fluxes are the constant coefficients $\omega^P_{MN}$, 
antisymmetric in the lower indices. In general, the twisted torus
has isometries with generators $Z_M$. The $\omega^P_{MN}$
turn out to be the structure constants of the Lie
algebra generated by the $Z_M$, i.e.  
\beq [Z_M, Z_N] = \omega^P_{MN}
Z_P \ .
\label{zal}
\eeq
Either from the Jacobi identity of the algebra or from the
Bianchi identity of (\ref{ocon}) one finds that the metric fluxes must
satisfy
\beq \omega^P_{[MN} \omega^S_{R]P} = 0 \ .
\label{jac}
\eeq
It can further be shown that $\omega^P_{PN}=0$ \cite{ss}.
The metric fluxes must be quantized by consistency of the twisted
torus structure \cite{lu}.

An useful result in the following is that we can 
contract the metric fluxes with a $p$-form $\cx$ to obtain 
a $(p+1)$-form $\omega \cx$ with components
\beq
(\omega \cx)_{LMN_1 \cdots N_{p-1}} = \omega^A_{[LM} \cx_{N_1 \cdots N_{p-1} ]}{}_A   \  .
\label{omX}
\eeq
Actually, for a constant form, $\omega \cx$ is basically $d\cx$ computed in the 
twisted torus.

The metric fluxes are even under the orientifold involution. Then they can be
of type $\omega^i_{ab}$, $\omega^i_{jk}$, $\omega^a_{ib}$, $i=1,2,3$,
$a=4,5,6$. As in the case of RR and NS fluxes, we only switch on metric fluxes
with one leg on each sub-torus. Thus, there are twelve free parameters
for which we use the notation
\beq
\bmat{c} \om^1_{56} \\ \om^2_{64} \\ \om^3_{45} \emat =
\bmat{c} a_1 \\ a_2 \\ a_3 \emat 
\quad ; \quad 
\bmat{ccc} \! \! \!
-\om^1_{23} & \, \om^4_{53} & \, \om^4_{26} \\ \, \om^5_{34} & \! \!
\! -\om^2_{31} & \, \om^5_{61} \\ \, \om^6_{42} & \, \om^6_{15} & \!
\! \! -\om^3_{12} \emat =
\bmat{ccc} b_{11} & b_{12} & b_{13} \\ b_{21} & b_{22}
& b_{23} \\ b_{31} & b_{32} & b_{33} \emat 
\ .
\label{abmatrix}
\eeq
The Jacobi identities imply the twelve constraints 
\beqa 
b_{ij}
a_j + b_{jj} a_i & = & 0 \quad ; \quad i \not= j \nonumber \\[0.2cm]
b_{ik} b_{kj} + b_{kk} b_{ij} & = & 0 \quad ; \quad i \not= j \not= k
\ .
\label{jacb} 
\eeqa 
Finally, the NS flux must satisfy the Bianchi identity \cite{km, vz}
\beq
\omega \ov{H}_3 = 0 \ ,
\label{omH}
\eeq
with the contraction defined in (\ref{omX}).
This constraint is satisfied automatically by the
particular fluxes (\ref{hfluxa}) and (\ref{abmatrix}). 


The superpotential induced by the fluxes can be obtained
by performing the explicit Kaluza-Klein reduction \cite{gl, vz}.
The RR fluxes generate a superpotential only for the K\"ahler moduli, namely
\beq 
W_K = 
\int_{\T^6} e^{J_c} \wedge \ov{F}_{RR} \ ,
\label{wkahler}
\eeq 
where $\ov{F}_{RR}$ represents a formal sum of the even RR
fluxes. NS and metric fluxes give a superpotential for
the dilaton and complex structure moduli that can be cast as
\beq 
W_Q= \int_{\T^6} \Omega_c \wedge (\ov{H}_3 + \omega J_c) \ .
\label{wqgen}
\eeq 
Recall that $\omega J_c$ is a 3-form as defined in (\ref{omX}).
Using previous results it is easy to compute $W_Q$. Combining with
$W_K$ yields the full superpotential 
\beqa 
W & = & e_0
+ ih_0 S + \sum_{i=1}^3 [(ie_i - a_i S - b_{ii}U_i -\sum_{j\not= i}
b_{ij}U_j)T_i - i h_iU_i] 
\nonumber \\[0.2cm] 
& - & q_1 T_2 T_3 -q_2 T_1 T_3 -q_3 T_1 T_2 + i m T_1 T_2 T_3 \ .
\label{wa}
\eeqa 
This result was first presented in \cite{vz} and analyzed in detail in
\cite{cfi}.

The fluxes also induce RR tadpoles. In this IIA orientifold there are
$C_7$ tadpoles. In fact, the ten-dimensional action has a piece
\beq
\int_{M_4 \times \T^6}[C_7 \wedge (m \ov{H}_3 + \omega \ov{F}_2 )] +
\sum_a N_a \int_{M_4 \times \Pi_a} C_7 \ .
\label{c7tad}
\eeq
The second term takes into account the coupling to O6-planes and stacks of
D6-branes wrapping factorizable 3-cycles
\beq
\Pi_a=(n_a^1, m_a^1)\otimes(n_a^2, m_a^2) \otimes(n_a^3, m_a^3) \  ,
\label{wns}
\eeq
and the corresponding orientifold images wrapping $\otimes_i (n_a^i, -m_a^i)$.
Here $n_a^i$ $(m_a^i)$ are the wrapping numbers along the $x^i$ $(y^i)$
torus directions. The O6-planes wrap $\otimes_i(1,0)$.
{}From the component of $C_7$ along $x^1$, $x^2$ and $x^3$ we obtain
\beq
\sum_a N_a n_a^1 n_a^2 n_a^3 + \oh(m h_0  +
a_1 q_1 + a_2 q_2 + a_3 q_3) = 16 \  .
\label{tadxxx}
\eeq
{}From other components of $C_7$ there are further cancellation
conditions
\beqa
\sum_a N_a n_a^1 m_a^2 m_a^3 + \oh (m h_1 - q_1b_{11} - q_2 b_{21} -
q_3 b_{31}) & = & 0 \ , \nonumber \\[0.2cm]
\sum_a N_a m_a^1 n_a^2
m_a^3 + \oh( m h_2 - q_1 b_{12} - q_2 b_{22} - q_3 b_{32}) & = & 0 \ ,
\label{tadodh} \\[0.2cm]
\sum_a N_a m_a^1 m_a^2 n_a^3 + \oh (m h_3 - q_1 b_{13} - q_2 b_{23} -
q_3 b_{33}) & = & 0 \ . \nonumber
\eeqa
In a $\Z_2 \times \Z_2$ set-up there are other O6-planes that
contribute -16 to the right hand side \cite{blt, cu}.

\subsection{IIB orientifold with O3-planes}
\label{sseciib}

We now discuss the IIB orientifold on $\T^6/[\Omega_P (-1)^{F_L} \sigma_B]$,
with involution acting on the K\"ahler
form and the holomorphic 3-form as $\sigma_B(J)=J$ and
$\sigma_B(\Omega)=-\Omega$. In terms of the coordinates  
this is $\sigma_B(x^i)=-x^i$ and $\sigma_B(y^i)=-y^i$.
Thus, there are O3-planes that span space-time.
Upon T-duality along $x^1$, $x^2$ and $x^3$, we recover the
IIA orientifold with O6-branes of the previous section.

We have  again seven diagonal closed moduli, the dilaton,  three K\"ahler moduli 
and three complex structure moduli. We denote them as $S$, $T_i$ and $U_i$,
even though they have different realizations in terms of the ten-dimensional 
degrees of freedom. In fact, the IIB and IIA moduli are related by T-duality
as $T_i \leftrightarrow U_i$, whereas $S$ is invariant. The IIB complex
structure fields are given directly by the toroidal complex structures,
i.e. $U_j = \tau_j$. The complex dilaton is instead
\beq
S= e^{-\phi} + i C_0   \ ,
\label{sdef}
\eeq
where $C_0$ is the R-R 0-form. The K\"ahler moduli can be extracted from the
complexified 4-form
\beq
\cj_c = C_4 + \frac{i}2 e^{-\phi} J \wedge J =  i\sum_{i=1}^3 T_i \wt{\om}_i  \  ,
\label{j2def}
\eeq
where $C_4$ is the RR 4-form.
The K\"ahler potential for the moduli has the same expression (\ref{kpot}).
The flux generated superpotential will be presented shortly.

The RR 3-form flux is odd under the orientifold involution.
The most general flux can then be written as
\beq
\ov{\cf}_3  =  -m \a_0
-e_0\b_0 + \sum_{i=1}^3 (e_i\a_i - q_i \b_i) \ .
\label{ffluxb} 
\eeq
Observe that the flux coefficients are the same that appear
in the RR IIA fluxes, c.f. (\ref{ffluxa}). This is in
agreement with T-duality. For the RR field strengths the
Buscher rule states \cite{Buscher, Hassan}
\beq
F_{M N_1 \cdots N_p}\,  {\stackrel{\mathsmaller{\T_M}}{\longleftrightarrow}} 
\, F_{N_1 \cdots N_p} \ ,
\label{tdualrr}
\eeq
where $\T_M$ is T-duality in the $x^M$ direction.
Then, performing T-dualities in $x^1, x^2, x^3$,
in that order, on $\ov{\cf}_3$ we obtain the IIA fluxes given in
(\ref{ffluxa}). Additional  T-dualities in $x^4, x^5, x^6$,
give the fluxes in the IIB orientifold on $\T^6/\Omega_P$
that has O9-planes. These results are summarized in table \ref{rriib}.

\begin{table}[htb] \footnotesize
\renewcommand{\arraystretch}{1.25}
\begin{center}
\begin{tabular}{cccc}
IIB/O3 & IIA/O6 & IIB/O9 & flux \\
\hline
$\ov{\cf}_{123}$ & $F_0$ & $\! \! \! -\bF_{456}$ & $-m$ \\
$\ov{\cf}_{423}$ & $F_{14}$ & $\bF_{156}$ & $-q_1$ \\
$\ov{\cf}_{153}$ & $F_{25}$ & $\bF_{426}$ & $-q_2$ \\
$\ov{\cf}_{126}$ & $F_{36}$ & $\bF_{453}$ & $-q_3$ \\
$\ov{\cf}_{156}$ & $F_{2536}$ & $\! \! \! -\bF_{423}$ & $\ \ e_1$ \\
$\ov{\cf}_{426}$ & $F_{1436}$ & $\! \! \! -\bF_{153}$ & $\ \ e_2$ \\
$\ov{\cf}_{453}$ & $F_{1425}$ & $\! \! \! -\bF_{126}$ & $\ \  e_3$ \\
$\ov{\cf}_{456}$ & $F_{142536}$ & $\bF_{123}$ & $-e_0$ \\
\end{tabular}
\end{center}
\caption{\small RR IIB/O3 fluxes and their T-duals.}
\label{rriib}
\end{table}

The NS 3-form flux is also odd under the orientifold involution.
We thus have the general expansion
\beq
\ov{\ch}_3  =  h_0\b_0 - \sum_{i=1}^3 a_i\a_i
+ \bar h_0\a_0 - \sum_{i=1}^3 \ba_i\b_i   \ .
\label{hfluxb}
\eeq
For NS fluxes we can apply Buscher rules \cite{Buscher} to T-dualize when 
$\ov{\ch}_3$ arises from a 2-form independent of
the dualized coordinates. In this case it is known that NS
fluxes can transform into metric fluxes \cite{glmw, kstt}. 
As reviewed in  \cite{cfi}, starting with
$\ov{\ch}_3  =  h_0\b_0 - \sum_i a_i\a_i$ and performing
T-dualities in $x^1, x^2, x^3$, leads to the IIA fluxes
\beq
\ov{H}_3= h_0 \b_0 \quad ; \quad  \omega^1_{56}= a_1 
\quad ; \quad  \omega^2_{64}= a_2 \quad ; \quad 
\omega^3_{45}= a_3 \ .
\label{nsiia}
\eeq
If $\ov{\ch}_3 \sim \a_0$, the 2-form would depend on one the $x^i$. 
If $\ov{\ch}_3 \sim \b_i$, one can still apply Buscher rules but 
they lead to more complicated geometries \cite{kstt}.
In \cite{stw} it was proposed that T-duality of the most general
NS flux will lead to metric as well as to {\em non-geometric} fluxes.
This will be the subject of next section.

In the IIB orientifold at hand there cannot be metric fluxes 
because the orientifold involution does not allow any even parameters
$\omega^P_{MN}$.

The NS and RR 3-form fluxes induce the well-known superpotential \cite{gvw}
\beq
\cw = \int_{\T^6} (\ov{\cf}_3 - iS \ov{\ch}_3) \wedge \Omega \ .
\label{wbfh}
\eeq
Substituting the fluxes we obtain
\beqa
\cw & = & e_0 + i\sum_{i=1}^3 e_i U_i
- q_1 U_2 U_3 - q_2 U_1 U_3 - q_3 U_1 U_2 +i m U_1 U_2 U_3
\nonumber \\[0.2cm]
& + & S \big[ ih_0 - \sum_{i=1}^3 a_i U_i
+i \ba_1 U_2 U_3 +i \ba_2 U_1 U_3 +i \ba_3 U_1 U_2 - \bar h_0 U_1 U_2 U_3 \big]
 \ .
\label{wnsrr}
\eeqa
To go to type IIA we just exchange $U_i \leftrightarrow T_i$. 
Comparing with (\ref{wa}) clearly shows that the two superpotentials
do not match. This hints at missing fluxes both in IIA and IIB. In the
next section we will see that after including the non-geometric fluxes
proposed in \cite{stw} the two superpotentials will be exact T-duals.

In this orientifold there is a $C_4$ tadpole induced by the fluxes. 
We know that it arises from the action term \cite{gkp}
\beq
\int_{M_4 \times \T^6} C_4 \wedge \ov{\ch}_3 \wedge \ov{\cf}_3 \ .
\label{c4tad}
\eeq
There are also contributions from O3-planes and a stack of $N_{\D3}$ D3-branes.
Substituting the fluxes and including
the sources we obtain the tadpole cancellation condition
\beq
N_{\D3} + \oh[m h_0 - e_0 \bh_0 + \sum_i (q_i
a_i + e_i \ba_i) ]  =  16 \ .
\label{d3o3}
\eeq
Upon T-duality this agrees with the $C_7$ tadpole (\ref{tadxxx}).
We also expect $C_8$ tadpoles to match (\ref{tadodh}) but they cannot
be induced by RR and NS fluxes alone, clearly some terms 
are missing. This is another indication
that non-geometric fluxes are required by T-duality.

\section{T-duality and non-geometric fluxes}
\label{secng}

There are two types of non-geometric fluxes introduced in \cite{stw},
the tensors $Q^{MN}_P$ and $R^{MNP}$ that are completely antisymmetric
in the upper indices. The $Q$'s are odd under the orientifold involution, while
the $R$'s are even. Recall that $\ov{H}_{MNP}$ is odd and $\om^M_{NP}$ is even. 
Now, the crucial property is that all these fluxes are related
by T-duality according to the chain
\beq
-\ov{H}_{MNP} \ {\stackrel{\T_M} {\longleftrightarrow}} \ \om^M_{NP}  \ 
{\stackrel{\T_N}{\longleftrightarrow}}  \ Q^{MN}_P  \ 
{\stackrel{\T_P}{\longleftrightarrow}}  \  -R^{MNP} \ .
\label{tdualns}
\eeq
We have introduced some extra signs in order to agree with the conventions
used in \cite{cfi}. For example, we have seen that $\ov{H}_{156}=-a_1$ 
transforms under $\T_1$-duality into $\om^1_{56}=a_1$. 

In IIB with O3-planes there are neither metric fluxes 
nor non-geometric fluxes of type $R$ because there are no such tensors even 
under the orientifold involution.
There are only odd non-geometric fluxes, denoted $\cq^{MN}_P$, that
comprise twenty-four free parameters
taking each index in a different sub-torus. There are also odd 
NS fluxes $\ov{\ch}$.

In IIA with O6-planes there are non-geometric fluxes $Q$ and $R$,
as well as NS $\ov{H}$ and metric fluxes $\omega$.
It is also interesting to consider
IIB with O9-planes in which the orientifold involution is the
identity. In this case there can only be even fluxes denoted
$\bom$ and $\bR$.  

Using the rule (\ref{tdualns}) and
starting with the NS and metric fluxes in the IIA side we can 
perform a chain of $\T_1$, $\T_2$ and $\T_3$ dualities,
i.e. mirror symmetry, to obtain the corresponding
fluxes in IIB. For example, under mirror symmetry, $\ov{H}_{ijc} \to - \cq_c^{ij}$, 
$\omega^i_{jk} \to  \cq_i^{jk}$, and so on. The results are summarized in table
\ref{nongeoiib}. Notice that the indices are ordered cyclically according to the 
sub-torus to which they belong.
The IIB/O9 fluxes are obtained from IIB/O3 by performing
six T-dualities, $\T_1, \cdots, \T_6$, or obviously from IIA/O6 by
applying only $\T_4, \T_5,  \T_6$. 

\begin{table}[htb] \footnotesize
\renewcommand{\arraystretch}{1.25}
\begin{center}
\begin{tabular}{cccc}
IIB/O3 & IIA/O6 & IIB/O9 & flux \\
\hline
$\bmat{ccc} \! \! \! \cq^{23}_4 & \cq^{31}_5 & \cq^{12}_6  \! \! \!\emat$ & 
$ \! \! -\bmat{ccc} \! \! \!  \ov{H}_{423} & \ov{H}_{153} & \ov{H}_{126}  \! \! \! \emat$ & 
$\bmat{ccc} \! \! \! \bom^4_{23} & \bom^5_{31} & \bom^6_{12}  \! \! \! \emat$ &
$ \! \! -\bmat{ccc} \! \! \!  h_1 & h_2 & h_3  \! \! \!\emat$  \\[0.35cm]
$\bmat{ccc} \! \! \!
-\cq^{23}_1 & \, \cq^{34}_5 & \, \cq^{42}_6 \\
\, \cq^{53}_4 & \! \! \! -\cq^{31}_2 & \, \cq^{15}_6 \\
\, \cq^{26}_4 & \, \cq^{61}_5 & \! \! \! -\cq^{12}_3 \emat $  &
$\bmat{ccc} \! \! \!
-\om^1_{23} & \, \om^4_{53} & \, \om^4_{26} \\ 
\, \om^5_{34} & \! \! \! -\om^2_{31} & \, \om^5_{61} \\ 
\, \om^6_{42} & \, \om^6_{15} & \! \! \! -\om^3_{12} \emat$ & 
$\bmat{ccc} \! \! \!
-\bom^1_{23} & \, \bom^5_{34} & \, \bom^6_{42} \\ 
\, \bom^4_{53} & \! \!\! -\bom^2_{31} & \, \bom^6_{15} \\ 
\, \bom^4_{26} & \, \bom^5_{61} & \!\! \! -\bom^3_{12} \emat$ &
$\bmat{ccc} b_{11} & b_{12} & b_{13} \\ 
b_{21} & b_{22} & b_{23} \\ 
b_{31} & b_{32} & b_{33} \emat$ \\
& & & \\[-0.4cm]
$\bmat{ccc} \! \! \! \cq^{56}_1 & \cq^{64}_2 & \cq^{45}_3  \! \! \! \emat$ & 
$ \! \! -\bmat{ccc} \! \! \!  R^{156} & R^{426} & R^{453}  \! \! \! \emat$ & 
$\bmat{ccc} \! \! \! \bom^1_{56} & \bom^2_{64} & \bom^3_{45}  \! \! \! \emat$ &
$ \! \! -\bmat{ccc} \! \! \!  \bh_1 & \bh_2 & \bh_3  \! \! \! \emat$  \\[0.35cm]
$\bmat{ccc} \! \! \!
-\cq^{56}_4 & \, \cq^{61}_2 & \, \cq^{15}_3 \\
\, \cq^{26}_1 & \! \! \! -\cq^{64}_5 & \, \cq^{42}_3 \\
\, \cq^{53}_1 & \, \cq^{34}_2 & \! \! \! -\cq^{45}_6 \emat$ &
$\bmat{ccc} \! \! \!
-Q^{56}_4 & \, Q^{26}_1 & \, Q^{53}_1 \\ 
\, Q^{61}_2 & \! \! \! -Q^{64}_5 & \, Q^{34}_2 \\ 
\, Q^{15}_3 & \, Q^{42}_3 & \! \! \! -Q^{45}_6 \emat$ &
$\bmat{ccc} \! \! \!
-\bom^4_{56} & \, \bom^2_{61} & \, \bom^3_{15} \\ 
\, \bom^1_{26} & \! \! \! -\bom^5_{64} & \, \bom^3_{42} \\ 
\, \bom^1_{53} & \, \bom^2_{34} & \! \! \! -\bom^6_{45} \emat$ &
$\bmat{ccc} 
\bb_{11} & \bb_{12} & \bb_{13} \\
\bb_{21} & \bb_{22} & \bb_{23} \\
\bb_{31} & \bb_{32} & \bb_{33} \emat$ 
\end{tabular}
\end{center}
\caption{\small Non-geometric IIB/O3 fluxes and their T-duals.}
\label{nongeoiib}
\end{table}

So far we have accounted for twelve non-geometric IIB/O3 fluxes,
those related to the IIA/O6 backgrounds $h_i$ (NS) and $b_{ij}$ (metric).
The IIB/O3 orientifold projection still allows another twelve
components for $\cq$, denoted $\bh_i$ and $\bb_{ij}$ as shown in
table \ref{nongeoiib}. Applying T-dualities we then obtain the 
corresponding fluxes in IIA/O6 and IIB/O9. 
These results are also displayed in table \ref{nongeoiib}.  

Finally, applying T-duality to the IIB/O3 NS fluxes
reveals some non-geometric fluxes in IIA/O6 and IIB/O9.
For example, acting 
with the chain of $\T_1$, $\T_2$ and $\T_3$ dualities gives 
$\ov{\ch}_{123} \to R^{123}$, $\ov{\ch}_{423} \to - Q_4^{23}$, etc..
We have already seen that the $\ov{\ch}_{ibc}$ are T-dual to IIA metric
fluxes. These results are collected in table \ref{nsiib}.
We now have a complete explicit dictionary to translate from
one orientifold to another.

\begin{table}[htb] \footnotesize
\renewcommand{\arraystretch}{1.25}
\begin{center}
\begin{tabular}{cccc}
IIB/O3 & IIA/O6 & IIB/O9 & flux \\
\hline
$\ov{\ch}_{123}$ & $R^{123}$ & $\bR^{123}$ & $\ \ \bh_0$ \\
$\ov{\ch}_{423}$ & $-Q_4^{23}$ & $\bR^{423}$ & $-\ba_1$ \\
$\ov{\ch}_{153}$ & $-Q_5^{31}$ & $\bR^{153}$ & $-\ba_2$ \\
$\ov{\ch}_{126}$ & $-Q_6^{12}$ & $\bR^{126}$ & $-\ba_3$ \\
$\ov{\ch}_{156}$ & $-\om^1_{56}$ & $\bR^{156}$ & $-a_1$ \\
$\ov{\ch}_{426}$ & $-\om^2_{64}$ & $\bR^{426}$ & $-a_2$ \\
$\ov{\ch}_{453}$ & $-\om^3_{45}$ & $\bR^{453}$ & $-a_3$ \\
$\ov{\ch}_{456}$ & $\ov{H}_{456}$ & $\bR^{456}$ & $\ \ h_0$ \\
\end{tabular}
\end{center}
\caption{\small NS IIB/O3 fluxes and their T-duals.}
\label{nsiib}
\end{table}

The next task is to determine the superpotential and tadpoles induced
by the non-geometric fluxes.
In sections \ref{ssecbng} and \ref{ssecang}
we consider type IIB with O3-planes and type IIA with O6-planes
to some extent. 
We will give explicit expressions for the superpotentials
as integrals involving the flux tensors and the complexified forms
that encode the moduli. Tadpoles of RR $C_p$ forms are written
in terms of the flux combinations that couple to them.
The IIB orientifold with O9-planes will be briefly surveyed.
The NS, metric and non-geometric fluxes are expected to satisfy
Bianchi identities that generalize (\ref{jac}) and (\ref{omH}). This
type of constraints will be derived in section \ref{ssecbianching}.
 
With the T-dual superpotential available, the next step is
to analyze the moduli potential. In section \ref{ssecvacua1}
we will discuss some classes of vacua and compare with previous results.

\subsection{T-dual superpotential and tadpoles in IIB with O3-planes}
\label{ssecbng}

We want to determine the superpotential and tadpoles induced
by the $\cq$ fluxes. An useful result is that we can 
contract a $p$-form $\cx$ with $\cq$ to obtain 
a $(p-1)$-form $\cq \cx$ with components
\beq
(\cq \cx)_{LM_1 \cdots M_{p-2}} = \frac12 \cq^{AB}_{[L} \cx_{M_1 \cdots M_{p-2} ]}{}_{AB}   \  .
\label{cqX}
\eeq
This is analogous to the contraction with $\omega$ defined in (\ref{omX}).

Observing the IIA result (\ref{wa}) it is clear that the $\cq$ fluxes
must induce new terms linear in the $T_i$ and up to cubic order in 
the $U_i$. Such terms can be generated by adding to
$\cw$ a piece $\int \cq \cj_c \wedge \Omega$, where $\cj_c$ is the 4-form
that encodes the K\"ahler moduli, c.f. (\ref{j2def}), and $\cq\cj_c$ is a 3-form
according to (\ref{cqX}). The complete IIB superpotential is then
\beq
\cw = \int_{\T^6} (\ov{\cf}_3 - iS \ov{\ch}_3 +  \cq \cj_c) \wedge \Omega \ .
\label{wbfhq}
\eeq
Substituting the fluxes yields
\beqa
\cw & = & e_0 + i\sum_{i=1}^3 e_i U_i
- q_1 U_2 U_3 - q_2 U_1 U_3 - q_3 U_1 U_2 +i m U_1 U_2 U_3
\nonumber \\[0.2cm]
& + & S \big[ ih_0 - \sum_{i=1}^3 a_i U_i
+i \ba_1 U_2 U_3 + i \ba_2 U_1 U_3 + i \ba_3 U_1 U_2 - \bar h_0 U_1 U_2 U_3 \big]
\label{wbtdual}  \\[0.2cm]
& +  & \sum_{i=1}^3 T_i \big[ -ih_i  - \sum_{j=1}^3
U_j b_{ji}  + i U_2 U_3 \bb_{1i}  + i U_1 U_3 \bb_{2i} 
+ i U_1 U_2  \bb_{3i} +  U_1 U_2 U_3 \bh_i \big]   \ .
\nonumber
\eeqa
The $\cq$-induced terms are in the last row.

The general superpotential agrees with the proposal of \cite{stw}
if we assume a symmetry under exchange of the three sub-tori.
This amounts to setting $T_i=T$ and $U_i=U$
together with the choice of fluxes
\beqa
& & e_i=e \quad ; \quad  q_i = q \quad ; \quad a_i=a \quad ; \quad
\ba_i = \ba   \quad ; \quad   h_i=h \quad ; \quad  \bh_i = \bh_i  \ ;
\nonumber \\[0.2cm]
& & b_{ij}=b \ (i \not= j) \quad ; \quad b_{ii}=\b \quad ; \quad 
\bb_{ij}=\bb \ (i\not= j) \quad ; \quad \bb_{ii}=\bbet  \ .
 \label{isofluxes}
\eeqa
We have also taken $b_{ij}=b_{ji}$ and  $\bb_{ij}=\bb_{ji}$.
The superpotential then reduces to
\beqa
\cw & = & e_0 + 3i e U - 3q U^2  +i m U^3 
\nonumber \\[0.2cm]
& + & S \big[ ih_0 - 3 a U + 3i \ba U^2  - \bh_0 U^3 \big]
\\[0.2cm]
\label{wiso}
& + & 3T \big[ -ih - (2b+\b) U + i (2\bb + \bbet) U^2  + \bh U^3 \big] \ .
\nonumber
\eeqa
The fluxes that enter in the superpotential must satisfy some 
tadpole cancellation conditions and Bianchi constraints that
will be examined in the next sections. 

In this orientifold there is a $C_4$ tadpole already discussed in 
section \ref{sseciib}. 
We also expect $C_8$ tadpoles that can receive
contributions from D7-branes and O7-planes. 
The flux piece must be a 2-form
suitable to wedge with $C_8$. A natural candidate is
$\cq \ov{\cf}_3$, where the 2-form is computed according to (\ref{cqX}). The
proposal for the $C_8$ tadpole is just
\beq
-\int_{M_4 \times \T^6} C_8 \wedge \cq\ov{\cf}_3  \ .
\label{c8tad}
\eeq
The minus sign in front is needed to match the known IIA results 
when only NS and metric fluxes are present. 
There are three different tadpoles according to the components of $C_8$ 
that can couple to $\D 7_i$-branes. As usual, a $\D7_i$-brane is 
transverse to $\T_i^2$ while wrapping $\T_j^2$ and
$\T_k^2$, $i\not= j\not= k$.
For example, the flux contribution to the $\D 7_1$ tadpole comes from
\beq
(\cq\ov{\cf}_3)_{14} = -m h_1 + e_0 \bh_1 + \sum_i (q_i b_{i1} +
e_i \bb_{i1}) \  .
\label{suno}
\eeq
Taking into account a number $N_{\D7_i}$ of $\D 7_i$-branes, and the flux tadpoles arising 
from (\ref{c8tad}), gives the cancellation conditions
\beq
-N_{\D7_i}  +  \oh \big[m h_i - e_0 \bh_i -
\sum_j (q_j b_{ji} + e_j \bb_{ji}) \big]  =  0 \ .
\label{tad7} 
\eeq
We have not included O7-planes, absent in a setup without $\Z_2
\times \Z_2$ orbifolding. A new interesting feature is the dependence of 
the tadpoles on all RR fluxes.

\subsection{T-dual superpotential and tadpoles in IIA with O6-planes}
\label{ssecang}

In this case there are non-geometric $Q$ and $R$ fluxes.
As in (\ref{cqX}), we can contract $Q$ with  a $p$-form $\cx$  
to obtain a $(p-1)$-form $Q \cx$. Analogously, contracting
with $R$ we obtain a $(p-3)$-form with components 
\beq
(R \cx)_{M_1 \cdots M_{p-3}} = \frac16 R^{ABC} \cx_{[M_1 \cdots M_{p-3} ]}{}_{ABC}   \  .
\label{RX}
\eeq
For example, the 3-form $R\ov{F}_6$ contributes to $C_7$ tadpoles. 

The $Q$ and $R$ fluxes are expected to induce superpotential terms quadratic
and cubic in the IIA K\"ahler moduli. There are appropriate 2 and 3-forms
that encode the required combination of the $T_i$, namely
\beqa
J_c^{(2)} & \equiv & \ts{\oh} J_c \wedge J_c = -T_2T_3 \, \wt{\om}_1
-T_1T_3 \, \wt{\om}_2 -T_1T_2 \, \wt{\om}_3
\nonumber \\[0.2cm]
J_c^{(3)} & \equiv & \ts{\frac16} J_c \wedge J_c \wedge J_c=
-iT_1T_2T_3 \, \a_0 \wedge \b_0  \  .
\label{jc23}
\eeqa
Then, the IIA superpotential T-dual to (\ref{wbfhq}) can be written as
\beq 
W = \int_{\T^6} \big[e^{J_c} \wedge \ov{F}_{RR} + 
\Omega_c \wedge (\ov{H}_3 + \omega J_c + QJ_c^{(2)} + RJ_c^{(3)}) \big] \ .
\label{wafhoqr}
\eeq
Substituting the fluxes precisely reproduces (\ref{wbtdual})
upon exchanging $T_i \leftrightarrow U_i$.

The idea behind the general formula for $W$ is to wedge
$\Omega_c$ with all available 3-forms.
An analogous reasoning suggests that the $C_7$ tadpoles due to
all fluxes follow from
\beq
\int_{M_4 \times \T^6} C_7 \wedge (-\ov{H}_3 \ov{F}_0 + \omega \ov{F}_2
- Q \ov{F}_4 + R \ov{F}_6)  \ .
\label{c7tadfull}
\eeq
The signs have been chosen to match results in type IIB. 
Including tadpoles due to O6-planes and stacks of intersecting D6-branes 
leads to the general cancellation conditions
\beqa
\sum_a N_a n_a^1 n_a^2 n_a^3 + \oh\big[m h_0 - e_0 \bh_0
+ \sum_i (q_i a_i + e_i \ba_i) \big] & = &  16 \  ,
\nonumber \\[0.2cm]
\sum_a N_a n_a^1 m_a^2 m_a^3 + \oh \big[m h_1 - e_0 \bh_1 -
\sum_i (q_i b_{i1} + e_i \bb_{i1})\big]  & = & 0 \ ,
\nonumber \\[0.2cm]
\sum_a N_a m_a^1 n_a^2 m_a^3 +
\oh\big[ m h_2 - e_0\bh_2 - \sum_i (q_i b_{i2} + e_i \bb_{i2}) \big] & = & 0 \ ,
\label{tadall} \\[0.2cm]
\sum_a N_a m_a^1 m_a^2 n_a^3 +
\oh \big[m h_3 - e_0 \bh_3 - \sum_i (q_i b_{i3} + e_i \bb_{i3}) \big] & = & 0 \ .
\nonumber
\eeqa
These agree with (\ref{d3o3}) and (\ref{tad7}).

\subsection{T-dual superpotential and tadpoles in IIB with O9-planes}
\label{sseciib9}

In this case the orientifold action is only $\Omega_P$. Since
the orientifold involution is the identity only even fluxes are allowed.
There are eight RR $\bF_{LMN}$, twenty-four metric $\bom^L_{MN}$, and eight 
non-geometric $\bR^{LMN}$. 
The components are displayed in tables \ref{rriib},  \ref{nongeoiib} and \ref{nsiib}.

The superpotential can be derived from IIB/O3 results by
implementing T-dualities in each of the six internal coordinates.
The moduli then transform as $S \leftrightarrow S$,    
$T_i \leftrightarrow T_i$, but  $U_i \leftrightarrow 1/U_i$.
The K\"ahler potential transforms as 
\beq
K \to K+\log|U_1U_2U_3|^2 \ .
\label{kudual}
\eeq
Invariance of the K\"ahler function, $\cg=K+ \log|\cw|^2$, then requires
\beq
W_{\rm O9} = \frac{-i\cw}{U_1U_2U_3} \ .
\label{wo9}
\eeq    
where we have chosen a convenient phase. 
Therefore, in terms of IIB/O9 moduli, 
\beqa
W_{\rm O9}  & = & m + i\sum_{i=1}^3 q_i U_i
+ e_1 U_2 U_3 + e_2 U_1 U_3 + e_3 U_1 U_2 - i e_0 U_1 U_2 U_3
\nonumber \\[0.2cm]
& + & S \big[ i\bh_0 + \sum_{i=1}^3 \ba_i U_i
+i a_1 U_2 U_3 + i a_2 U_1 U_3 + i a_3 U_1 U_2 + h_0 U_1 U_2 U_3 \big]
\label{wo9tdual}  \\[0.2cm]
& +  & \sum_{i=1}^3 T_i \big[ -i\bh_i  + \sum_{j=1}^3
\bb_{ji} U_j  + i b_{1i}  U_2 U_3  + i b_{2i}  U_1 U_3 
+ i b_{3i} U_1 U_2 -  h_i U_1 U_2 U_3  \big]   \ .
\nonumber
\eeqa
In absence of metric $\bom$ and non-geometric $\bR$
fluxes $W_{\rm O9}$ depends only on the complex structure moduli.
Linear terms in $T_i$ and $S$ are induced by $\bom$ and $\bR$
respectively.

Now there are O9-planes and we can add D9-branes.
We then anticipate that fluxes contribute to a $C_{10}$ tadpole.
Indeed, there is a candidate tadpole term
\beq
\int_{M_4 \times \T^6} C_{10} \wedge \bR\bF_3  \ ,
\label{c10tad}
\eeq 
where $\bR\bF_3$ is a 0-form according to (\ref{RX}).
Substituting the fluxes and including sources gives 
\beq
N_{\D9} + \oh[m h_0 - e_0 \bh_0 + \sum_i (q_i
a_i + e_i \ba_i) ]  =  16 \ .
\label{d9o9}
\eeq
To match the IIB/O3 results there must also be $C_6$ tadpoles.
With the available fluxes we can indeed have a term
\beq
\int_{M_4 \times \T^6} C_6 \wedge \bom\bF_3  \ ,
\label{c6tad}
\eeq
where $\bom\bF_3$ is a 4-form according to (\ref{omX}).
We can also add $\D 5_i$-branes that wrap $\T_i^2$.
We then find cancellation conditions
\beq
N_{\D5_i}  +  \oh \big[m h_i - e_0 \bh_i -
\sum_j (q_j b_{ji} + e_j \bb_{ji}) \big]  =  0 \ .
\label{tad5} 
\eeq
We have not included O5-planes.

\subsection{Constraints on NS, metric and non-geometric fluxes}
\label{ssecbianching}

We saw in chapter 2 that the geometric fluxes
$\omega^M_{NP}$ are subject to the Bianchi identities in eq.(\ref{jac}).
To find the analogous constraints for non-geometric fluxes 
we will follow the approach of \cite{stw}, see also \cite{dh2}.
The strategy is to extend the algebra of isometry generators
$Z_M$ to include generators $X^M$, $M= 1,\cdots, 6$. 
The $X^M$ are associated to gauge symmetries arising from
reduction of the $B$-field on $\T^6$ with fluxes \cite{km}.
The extended algebra has the NS, metric and non-geometric
fluxes as structure constants. The most general algebra is then 
\beqa
[Z_M, Z_N] & = &  -\ov{H}_{MNP} X^P + \omega^P_{MN} Z_P   \  ,
\nonumber \\[0.2cm]
[Z_M, X^P] & = &  -\omega^P_{MN} X^N + Q^{PR}_M  Z_R    \  ,
\label{algxz} \\[0.2cm]
[X^M, X^N] & = &   Q^{MN}_P X^P - R^{MNP} Z_P    \  .
\nonumber 
\eeqa
The Jacobi identities of the algebra give constraints on the fluxes\footnote{
In contrast to the algebras considered in \cite{DallAgata}, here there are 
always six  $Z_M$ and six $X^M$ generators, as required to account for all fluxes.}.

The proposed algebra actually applies to any of the IIA or IIB 
orientifolds, provided that all fluxes allowed by the orientifold
action are kept in each case. In fact, using (\ref{tdualns}) and 
\beq
Z_M \, {\stackrel{\T_M} {\longleftrightarrow}} \, X^M  \ ,
\label{zxdual}
\eeq
we see that the algebra is invariant under $\T_M$-duality.
Written in terms of the various tensors, the
Jacobi identities take a different form in each case. However, in terms of
the individual flux parameters that appear in the T-dual superpotential there
is just one set of constraints valid on all orientifolds. 

It is convenient to work with the IIB with O3-planes in which only 
NS $\ov{\ch}$ and non-geometric $\cq$ fluxes appear. The $ZZZ$ Jacobi identity
leads to
\beq
\cq^{RP}_{[L} \ov{\ch}_{MN]P} =  0  \ . 
\label{jacqh} 
\eeq
Substituting the fluxes in tables \ref{nongeoiib} and \ref{nsiib} then
yields
\beqa
\bar{h}_0h_j+\bar{a}_ib_{ij}+\bar{a}_jb_{jj}-a_k\bar{b}_{kj}&=&0  \ ,
\label{hq1}\\[0.2cm]
h_0\bar{h}_j+a_i\bar{b}_{ij}+a_j\bar{b}_{jj}-\bar{a}_kb_{kj}&=&0 \  ,
\label{hq2}\\[0.2cm]
\bar{h}_0b_{kj}+\bar{a}_i\bar{b}_{jj}+\bar{a}_j\bar{b}_{ij}-a_k\bar{h}_j&=&0 \ ,
\label{hq3}\\[0.2cm]
h_0\bar{b}_{kj}+a_ib_{jj}+a_jb_{ij}-\bar{a}_kh_j&=&0   \ .
\label{hq4}
\eeqa
In all cases $i\not=j\not=k$.
The $XXX$ Jacobi identity simply gives
\beq
\cq_P^{[MN} \cq_R^{L]P} =  0 \ . 
\label{jacqq} 
\eeq
In terms of the explicit fluxes
\beqa
-b_{ii}b_{jk}+\bar{b}_{ki}h_k+h_i\bar{b}_{kk}-b_{ji}b_{ik}&=&0  \ , 
\label{qq1}\\[0.2cm]
-\bar{b}_{ii}\bar{b}_{jk}+b_{ki}\bar{h}_k+\bar{h}_ib_{kk}-\bar{b}_{ji} \bb_{ik}&=&0  \ , 
\label{qq2}\\[0.2cm]
-b_{ii}\bar{b}_{ij}+\bar{b}_{ji}b_{jj}+h_i\bar{h}_j-b_{ki}\bar{b}_{kj}&=&0  \ , 
\label{qq3}\\[0.2cm]
\bar{b}_{ii}b_{ij}-b_{ji}\bar{b}_{jj}+h_i\bar{h}_j-b_{ki}\bar{b}_{kj}&=&0  \  .
\label{qq4}
\eeqa
In all cases $i\not=j\not=k$.
There are no further constraints from other Jacobi identities.

With the isotropic fluxes given in (\ref{isofluxes}) the constraints read
\beqa
\bh_0 h +\ba(b + \b)-a\bb &=&0 \ , 
\label{ihq1}\\[0.2cm]
h_0 \bh +a (\bb + \bbet) -\ba b &=&0  \ , 
\label{ihq2}\\[0.2cm]
\bh_0 b + \ba(\bb +\bbet) -a \bh &=&0  \ , 
\label{ihq3}\\[0.2cm]
h_0\bb + a (b + \b) - \ba h &=&0   \ , 
\label{ihq4}\\[0.4cm]
h(\bb + \bbet) - b(b +\b) &=&0   \ , 
\label{iqq1}\\[0.2cm]
\bh(b + \b) - \bb(\bb +\bbet) &=&0  \ , 
\label{iqq2}\\[0.2cm]  
h \bh  - b \bb &=&0 \ .
\label{iqq34}
\eeqa
Some classes of solutions are:
\beqa
&1.& h=\bh=b=\b=\bb=\bbet=0 \ (\cq=0) \quad ; \quad
a, \ba, h_0, \bh_0\not=0  \ (\ov{\ch}\not=0)  \  .
\label{ohf} \\[0.4cm] 
&2.& a=\ba=h_0=\bh_0=0  \ (\ov{\ch}=0)  
\quad ; \quad
h \bh = b \bb  \quad ; \quad h(\bb + \bbet) = b(b+\b)  \  .
 \label{hfz} \\[0.4cm]
&3.& a=\ba=h=\bh=b=\bb=0  
\quad ; \quad
h_0, \bh_0, \b, \bbet \not=0  \  .
\label{bpnz} \\[0.4cm] 
&4.& \b=-b \quad ; \quad \bbet=-\bb \quad ; \quad 
h \bh = b \bb  \quad ; \quad \bh_0 h = a \bb \quad ; \quad h_0 \bh = \ba b   \  .
\label{nadahqz}
\eeqa

\subsection{Some vacua with T-dual fluxes}
\label{ssecvacua1}

In this section we work in the IIA/O6 setup for
ease of comparison with results of \cite{cfi}.
Our purpose is to see the effect of the new non-geometric
fluxes in some simple examples. We focus on no-scale
type of superpotentials depending only on four moduli.
In this case the scalar potential is positive definite
and is minimized with respect to all fields when
the four covariant derivatives of $W$ vanish.

To be concrete we consider the moduli $S$, $U_1$,
$T_2$ and $T_3$. Turning on all fluxes visible to 
these fields gives the generic superpotential 
\beqa
W & = & e_0 + i e_2 T_2 + i e_3 T_3 - q_1 T_2 T_3
+ S \big[ ih_0 - a_2 T_2 - a_3 T_3  +i \ba_1 T_2 T_3 \big]
\nonumber \\[0.2cm]
& -  & U_1 \big[ ih_1  + b_{21} T_2 +  b_{31} T_3 - i\bb_{11} T_2 T_3 \big]   \ .
\label{exwa}
\eeqa
It is easily proven that the set of fluxes in $W$ satisfies the Bianchi
identities of the previous section. Moreover, these fluxes do not
contribute to tadpoles.

When one of the non-geometric flux parameters is zero
we can map $W$ to one of the cases studied in \cite{cfi}. For instance,
when $\ba_1=0$, redefining $U_1 \to T_1$, and relabelling
fluxes appropriately, brings us to the NS-3 example of \cite{cfi}.
If $a_2 = a_3=0$, we can compare with the simpler NS-1 model. In this case
we conclude that there are minima only if $h_0 \not=0$, $\bb_{11} \not=0$
and furthermore
\beq
h_1 \bb_{11} = b_{31} b_{21} \quad ; \quad
e_2 \bb_{11} = -q_1 b_{21} \quad ; \quad
e_3 \bb_{11} = -q_1 b_{31} \ .
\label{surprise}
\eeq
We also find that axions are fixed but the real parts of moduli remain
undetermined except for a relation $h_0 \re S = \bb_{11} \re U_1 \re T_2 \re T_3$.
The situation with $\bb_{11}=0$ also corresponds to the NS-3 example \cite{cfi}. 
It is only when $\ba_1 \not=0$ and  $\bb_{11} \not=0$, so that both
cubic terms are present in $W$, that we can have a different kind of
no-scale model. 

We have analyzed the case with both non-geometric fluxes turned on to some
extent. To simplify we choose $a_i=a$, $b_{i1}=b$ and $e_i=e$, which allows
vacua with $T_2=T_3=T$. Some generic results can be extracted.
For instance,
\beqa
\ba\, \im S + \bb\, \im U & = & -q \ ,
\nonumber \\
(a + \ba \, \im T)\, \re S & = &  (b + \bb \, \im T)\, \re U  \ , 
\label{genns}
\eeqa
where we have dropped subindices. There are more equations to be solved. 
In general only the ratio of $\re S$ and $\re U$ is fixed.
When $a\bb \not= b \ba$, the fluxes generically determine all axions as well as
$\re T$. When $a\bb = b \ba$, and $e\ba = -q a$ by consistency,
there are two types of solutions. In one type, with $hq \not=-eb$,
all axions are determined, in fact $\im T=-a/\ba$, and $(\re T)^4$ is completely 
fixed in terms of fluxes alone. In the other type, with fluxes further satisfying
$h e = - e_0 b$ and $h_0 b = -h a$, all axions are undetermined
and $(\re T)^2$ can only be given in terms of $\im T$.  
For instance by choosing the fluxes $a=\ba=b=\bb=h_0=e_0=e=-q=2$
and $h=0$ we find the solution
\beq
\im S =-\im T=1 \quad ;   \quad  \im U =0  \quad ;  \quad 
\re T =2^{\frac14} \quad ;   \quad
\re U=\sqrt{2}\, \re S  \ .
\label{NSsol2}
\eeq
Taking larger fluxes it should be possible to obtain
larger $\re T$.

To summarize, adding non-geometric backgrounds leads to
new no-scale vacua, inequivalent to those with only RR, NS
and metric fluxes. However, the examples that we have examined 
have qualitative properties analogous to the vacua analyzed in \cite{cfi}.
In section \ref{secvacua} we will briefly consider Minkowski
vacua in presence of non-geometric fluxes.

\section{IIB S-duality and fluxes}

We know that type IIB string theory is S-duality invariant. Upon
the orientifold compactification here considered we still expect the theory 
to reflect this underlying invariance. In fact, in the absence of 
non-geometric fluxes, when only  $\ov{\ch}_3$ and $\ov{\cf}_3$ are present,
the theory is explicitly S-duality invariant because these fluxes 
transform appropriately. On the other hand, once we have added 
non-geometric fluxes $\cq$ the theory does not respect S-duality.
We will see in this chapter that S-duality invariance requires the 
presence of extra flux degrees of freedom $\cp$. 

We want to implement invariance under the $SL(2,\Z)$ S-duality transformations
\beq
S \to \frac{kS -i\ell}{imS+n} \quad ; \quad kn-\ell m=1
\quad ; \quad k,\, \ell, \, m, \, n \in \Z  \ .
\label{sl2z}
\eeq
The factors of $i$ are needed since in our conventions $\re S =
1/g_s$. The K\"ahler potential, $K=-\log(S+S^*) + \cdots$,
transforms as
\beq
K \to K+\log|imS+n|^2 \ .
\label{ksdual}
\eeq
Thus, the K\"ahler function, $\cg=K+ \log|\cw|^2$, is invariant
provided the superpotential verifies
\beq
\cw \to \frac{\cw}{imS+n} \ .
\label{wsdual}
\eeq
With only NS and RR fluxes turned on this follows
simply because under S-duality the NS and RR 3-forms transform as
\beq
\bmat{c} \ov{\cf}_3 \\ \ov{\ch}_3 \emat \to 
\bmat{cc} k & \ell \\ m & n \emat \!\! \bmat{c} \ov{\cf}_3 \\ \ov{\ch}_3
\emat \ .
\label{sl2zflux}
\eeq
In particular, when $S \to 1/S$, $\ov{\cf}_3 \to -\ov{\ch}_3$ and
$\ov{\ch}_3 \to \ov{\cf}_3$. 

The question is now how
to maintain S-duality after including the non-geometric fluxes $\cq$.
To obtain a full S-dual superpotential we simply propose
to add a new set of fluxes, denoted $\cp$, with the same tensor structure 
and number of components as $\cq$. 
Concretely, we conjecture that the superpotential is given by
\beq
\cw = \int_{\T^6} \big[(\ov{\cf}_3 - iS \ov{\ch}_3) + (\cq -iS
\cp)\cj_c \big] \wedge \Omega \ .
\label{wbfull}
\eeq
The action will be invariant as long as $\cq$ and $\cp$ fluxes
transform as 
\beq
\bmat{c} \cq \\ \cp \emat \to 
\bmat{cc} k & \ell \\ m & n \emat \!\! \bmat{c} \cq \\ \cp
\emat \ .
\label{pintoq}
\eeq
In particular, when $S \to 1/S$, one has $\cq \to -\cp$ and
$\cp \to \cq$.

The new objects $\cp^{MN}_P$ are some sort of RR non-geometric
fluxes. For the components we use the notation
\beq
\bmat{c} \cp^{23}_4 \\ \cp^{31}_5 \\ \cp^{12}_6 \emat =
\bmat{c} -f_1 \\ -f_2 \\ -f_3 \emat 
\quad ; \quad 
\bmat{ccc} \! \! \!
-\cp^{23}_1 & \, \cp^{34}_5 & \, \cp^{42}_6 \\
\, \cp^{53}_4 & \! \! \! -\cp^{31}_2 & \, \cp^{15}_6 \\
\, \cp^{26}_4 & \, \cp^{61}_5 & \! \! \! -\cp^{12}_3 \emat = 
\bmat{ccc} g_{11} & g_{12} & g_{13} \\
g_{21} & g_{22} & g_{23} \\
g_{31} & g_{32} & g_{33} \emat 
\ ,
\label{pfluxes1}
\eeq
together with
\beq
\bmat{c} \cp^{56}_1 \\ \cp^{64}_2 \\ \cp^{45}_3 \emat =
\bmat{c} -\ff_1 \\ -\ff_2 \\ -\ff_3 \emat 
\quad ; \quad 
\bmat{ccc} \! \! \!
-\cp^{56}_4 & \, \cp^{61}_2 & \, \cp^{15}_3 \\
\, \cp^{26}_1 & \! \! \! -\cp^{64}_5 & \, \cp^{42}_3 \\
\, \cp^{53}_1 & \, \cp^{34}_2 & \! \! \! -\cp^{45}_6 \emat = 
\bmat{ccc} \bg_{11} & \bg_{12} & \bg_{13} \\
\bg_{21} & \bg_{22} & \bg_{23} \\
\bg_{31} & \bg_{32} & \bg_{33} \emat 
\ .
\label{pfluxes2}
\eeq
The superpotential generated by the $\cp$ fluxes alone is
\beqa
\cw_\cp  & = &  -S\sum_{i=1}^3 f_i T_i +iS
\sum_{i,j=1}^3 U_j g_{ji} T_i + S U_2 U_3 \sum_{i=1}^3 \bg_{1i}
T_i + S U_1 U_3 \sum_{i=1}^3 \bg_{2i} T_i
\nonumber \\[0.2cm]
& + & S U_1 U_2  \sum_{i=1}^3 \bg_{3i} T_i
 -iS    U_1 U_2 U_3 \sum_{i=1}^3 \ff_i T_i   \ .
\label{wpdetail}
\eeqa
We thus see that we get new superpotential 
couplings which are linear in $S$ and $T_i$ and  
up to cubic order in the $U_i$. 
The $\cp$ fluxes will also give rise to modifications 
to tadpole conditions and to 
Jacobi constraints involving these new fluxes. We now discuss
these issues in turn.

\subsection{S-dual tadpoles}
\label{ssectadsdual}

The $C_4$ tadpole term (\ref{c4tad}) is S-duality invariant  because $C_4$
is invariant whereas $\ov{\cf}_3$ and $\ov{\ch}_3$ transform as in (\ref{sl2zflux}).
On the contrary, the $C_8$ tadpole (\ref{c8tad}) is not S-duality invariant,
as one can easily check. 
In fact this is expected from the known fact \cite{mo, dlt, el} that 
the $C_8$ RR-form  is one component in a  $SL(2,\Z)$ triplet of 
8-forms, $(C_8,\widetilde{C}_8,C_8')$. Under $S\rightarrow 1/S$ they
transform as
\beqa
C_8 \ & \longrightarrow & \  - \widetilde{C}_8 \nonumber \ , \\
\widetilde{C}_8 \ & \longrightarrow & \  -  {C}_8  \ , \label{triplet} \\ 
C_8' \ & \longrightarrow & \  - {C}_8'   \ .
\nonumber
\eeqa
There is a constraint among the field strengths of these  8-forms
so that there are only two propagating degrees of freedom.
These three forms may be sourced by three types of 7-branes,
$\D7$-branes, ${\rm NS}7$-branes and certain other 7-branes called
I7  in  \cite{brkor}. In our factorized torus 
each of them will come in three varieties, $\D7_i$, ${\rm NS}7_i$ and ${\rm I}7_i$,
with $i=1,2,3$,  labelling one of the three tori transverse to the brane. 
Starting from eq.(\ref{c8tad})   and imposing $SL(2,\Z)_S$ invariance
of the action one arrives at
\beq
 \ \int_{M_4 \times \T^6}  -\,  C_8 \wedge \cq\ov{\cf}_3 \, 
+ \, \widetilde{C}_8 \wedge \cp\ov{\ch}_3 \,  + \, 
C_8' \wedge (\cq\ov{\ch}_3+\cp\ov{\cf}_3)  \ .
\label{tadc8}
\eeq
Just like $\cq\ov{\cf}_3$, here
$\cp\ov{\ch}_3$, $\cq\ov{\ch}_3$ and $\cp\ov{\cf}_3$ are 2-forms computed as 
in (\ref{cqX}). For example, 
$(\cq \ov{\ch})_{LM} = \frac12 \cq^{AB}_{[L} \ov{\ch}_{M]}{}_{AB}$.

The first two terms in eq.~(\ref{tadc8}) give rise to tadpoles of $\D 7_i$-branes and
their S-dual ${\rm NS}7_i$-branes. Looking at components we obtain the cancellation
conditions
\beqa
-N_{{\rm D}7_i} +  \oh \big[m h_i - e_0 \bh_i -
\sum_j (q_j b_{ji} + e_j \bb_{ji}) \big]  & = &  0 \ ,
\label{tad7j} \\[0.2cm]
-N_{{\rm NS}7_i} +  \oh \big[
h_0\ff_i  - \bh_0 f_i - \sum_j (\ba_j g_{ji} - a_j \bg_{ji})\big] & = & 0 \ .
\label{tadNSj}
\eeqa
Concerning the third term, one  observes that 
the flux combinations coupling to $C_8'$ do not have RR character. In fact  
they are rather related to NS Bianchi identities. 
In particular, in section \ref{ssecbianching} we found that 
$\cq^{AB}_{[L} \ov{\ch}_{MN]}{}_B=0$. It is easy to show
that for our class of fluxes this then implies $\cq\ov{\ch}_3=0$.
We discuss further this issue in the next section.

\subsection{S-dual Bianchi constraints}
\label{ssecbianchisdual}

In section \ref{ssecbianching} we discussed the Bianchi identities
leading to constraints on the $\ov{\ch}_3$ and $\cq$ fluxes.
By S-duality  we expect constraints on the RR fluxes $\ov{\cf}_3$ 
and the new fluxes $\cp$. 
To begin we consider the identity (\ref{jacqq}).
To achieve closure under $SL(2,\Z)_S$ we find that the
condition $\cq_P^{[MN} \cq_R^{L]P}=0$, schematically
$\cq \cdot \cq=0$, remains valid and that 
there are actually two new constraints.
The point is that $\cq \cdot \cq$ is a component in a triplet
of $SL(2,\Z)_S$. Acting with $S \to 1/S$ we find that another
component is $\cp \cdot \cp$, with corresponding Bianchi
identity given by
\beq
\cp_P^{[MN} \cp_R^{L]P} =  0 \ . 
\label{jacpp} 
\eeq
Finally, applying a translation $S \to (S +i)$ shows that the third
triplet component is $(\cq \cdot \cp + \cp \cdot \cq)$. Thus,
there is also a constraint
\beq
\cq_P^{[MN} \cp_R^{L]P} + \cp_P^{[MN} \cq_R^{L]P} =  0 \ . 
\label{jacqp} 
\eeq
Notice that the left hand side has net RR charge so that it is
potentially related to tadpoles. However, given the tensor structure,
it is not clear how it could couple to the known RR forms.
In absence of sources we are led to enforce the equality to zero.

In components, eq.~(\ref{jacpp}) breaks into ($i\not=j\not=k$)
\beqa
-g_{ii}g_{jk}+\bar{g}_{ki}f_k+f_i\bar{g}_{kk}-g_{ji}g_{ik}  &=&0   \ , 
\label{pp1}\\[0.2cm]
-\bar{g}_{ii}\bar{g}_{jk}+g_{ki}\bar{f}_k+\bar{f}_ig_{kk}-\bar{g}_{ji} \bg_{ik}  &=&0   \ , 
\label{pp2}\\[0.2cm]
-g_{ii}\bar{g}_{ij}+\bar{g}_{ji}g_{jj}+f_i\bar{f}_j-g_{ki}\bar{g}_{kj} &=&0   \ , 
\label{pp3}\\[0.2cm]
\bar{g}_{ii}g_{ij}-g_{ji}\bar{g}_{jj}+f_i\bar{f}_j-g_{ki}\bar{g}_{kj} &=&0   \ .
\label{pp4}
\eeqa
These are completely analogous to the conditions following from
$\cq_P^{[MN} \cq_R^{L]P}=0$, given in eqs.~(\ref{qq1})-(\ref{qq4}).
In the particular case in which one imposes 
a symmetry under exchange of the three sub-tori one has 
\beqa 
& & f_i= f \quad ; \quad \ff_i = \ff  \ ;
\label{giso}  \\ 
& & g_{ij}=g \ (i \not= j) \quad ; \quad g_{ii}=\g \quad ; \quad 
\bg_{ij}=\bg \ (i\not= j) \quad ; \quad \bg_{ii}=\bgam  \ .
\nonumber
\eeqa
We have further assumed  $g_{ij}=g_{ji}$ and  $\bg_{ij}=\bg_{ji}$.
In this case we find the simplified set of conditions
\beqa
f(\bg + \bgam) - g(g +\g) &=&0  \ , 
\label{ipp1}\\[0.2cm]
\ff(g + \g) - \bg(\bg +\bgam) &=&0   \ , 
\label{ipp2}\\[0.2cm]
f \ff  - g \bg &=&0   \ .
\label{ipp34}
\eeqa
A simple solution is $f=\ff=g=\bg=0$, but $\g, \bgam \not=0$. 
Another solution is $\g=-g$, $\bgam=-\bg$ and $\ff = g \bg/f$.

Concerning eq.~(\ref{jacqp}), it gives rise to the
four additional constraints ($i\not=j\not=k$)
\beqa
b_{kk}\bg_{kj} -h_k \ff_j - \bb_{jk}g_{jj} + b_{ik}\bg_{ij}      
+ g_{kk}\bb_{kj} - f_k \bh_j  - \bg_{jk}b_{jj} + g_{ik}\bb_{ij}  &=&0   \ , 
\label{qp1}\\[0.2cm]
b_{kk} g_{ij} -h_k \bg_{jj} - \bb_{jk} f_j + b_{ik} g_{kj} 
+ g_{kk} b_{ij} - f_k \bb_{jj}  - \bg_{jk} h_j + g_{ik} b_{kj}  &=&0   \ , 
\label{qp2}\\[0.2cm]
\bb_{kk} \bg_{ij} - \bh_k g_{jj} - b_{jk} \ff_j + \bb_{ik} \bg_{kj} 
+ \bg_{kk} \bb_{ij} - \ff_k b_{jj}  - g_{jk} \bh_j + \bg_{ik} \bb_{kj}  &=&0   \ , 
\label{qp3}\\[0.2cm]
\bb_{kk} g_{kj} - \bh_k f_j - b_{jk} \bg_{jj} + \bb_{ik} g_{ij}      
+ \bg_{kk} b_{kj} - \ff_k h_j  - g_{jk} \bb_{jj} + \bg_{ik} b_{ij}  &=&0   \ . 
\label{qp4}
\eeqa
For isotropic fluxes (\ref{isofluxes}) and (\ref{giso}) they reduce to
\beqa
b \bg + g \bb - h \ff - f\bh &=&0   \ , 
\label{iqp1}\\[0.2cm]
g(b+ \b) - h(\bg + \bgam) - f(\bb + \bbet) + b(g+ \g) &=& 0 \ , 
\label{iqp2}\\[0.2cm]
\bg(\bb+ \bbet) - \bh(g + \g) - \ff(b + \b) + \bb(\bg+ \bgam) &=& 0 \ . 
\label{iqp4}
\eeqa
The simple solution $f=\ff=g=\bg=0$ also
works, supplemented with $b\g = h \bgam$ and $\bb\bg = \bh \g$.

Let us now discuss the modifications to the identity
eq.~(\ref{jacqh}). In this case we can construct the
fully $SL(2,\Z)_S$ invariant condition 
\beq
\cq^{RP}_{[L} \ov{\ch}_{MN]P} - \cp^{RP}_{[L} \ov{\cf}_{MN]P} = 0 \  .  
\label{jacqhpf} 
\eeq
Substituting the flux components leads to ($i\not=j\not=k$)
\beqa
\bar{h}_0h_j + \bar{a}_i b_{ij} + \bar{a}_j b_{jj}-a_k\bar{b}_{kj}
+ mf_j - q_ig_{ij} - q_jg_{jj} - e_k\bar{g}_{kj} &=&0  \ , 
\label{hqfp1}\\[0.2cm]
h_0\bar{h}_j+a_i\bar{b}_{ij}+a_j\bar{b}_{jj}-\bar{a}_kb_{kj}
- e_0\bar{f}_j - e_i\bar{g}_{ij} - e_j\bar{g}_{jj} - q_kg_{kj} &=&0  \ , 
\label{hqfp2}\\[0.2cm]
\bar{h}_0b_{kj} + \bar{a}_i\bar{b}_{jj} + \bar{a}_j\bar{b}_{ij} - a_k\bar{h}_j
+ mg_{kj} - q_i\bar{g}_{jj} - q_j\bar{g}_{ij} - e_k\bar{f}_j  &=&0   \ , 
\label{hqfp3}\\[0.2cm]
h_0\bar{b}_{kj} + a_ib_{jj} +a_jb_{ij}-\bar{a}_kh_j
- e_0\bar{g}_{kj} - e_ig_{jj} - e_jg_{ij} - q_kf_j  &=&0   \ .
\label{hqfp4}
\eeqa
With the simple isotropic fluxes these constraints read
\beqa
\bh_0 h +\ba(b + \b)-a\bb 
+ m f - q (g + \g) - e \bg &=&0  \ , 
\label{ihqfp1}\\[0.2cm]
h_0 \bh +a (\bb + \bbet) -\ba b 
-e_0 \ff - e (\bg + \bgam) -qg &=&0  \ , 
\label{ihqfp2}\\[0.2cm]
\bh_0 b + \ba(\bb +\bbet) -a \bh 
+ m g - q (\bg + \bgam) - e\ff &=&0  \ , 
\label{ihqfp3}\\[0.2cm]
h_0\bb + a (b + \b) - \ba h 
-e_0 \bg - e (g + \g) -q f &=&0  \  .
\label{ihqfp4}
\eeqa
We will discuss some solutions in section \ref{secvacua}.

The combination $(\cq \cdot \ov{\ch}_3 - \cp \cdot \ov{\cf}_3)$
appearing in the identity (\ref{jacqhpf}) is an $SL(2,\Z)_S$ singlet. 
Now, with the doublets $(\cq, \cp)$ and $(\ov{\cf}_3, \ov{\ch}_3)$ 
we can also form a triplet which is rather
related to tadpoles of RR 8-forms as we saw in the previous section.
In particular, concerning the $C_8'$ flux tadpole in eq.~(\ref{tadc8}),
we see that in general it does not cancel because 
neither $\cq \cdot \ov{\ch}_3$  nor  $\cp \cdot \ov{\cf}_3$
has to vanish separately.
In fact, starting with the tadpole term of $C_8'$ 
we obtain the cancellation condition 
\beq
N_{{\rm I}7_i} +  \oh \big[
e_0\ff_i  - m f_i + \sum_j (q_j g_{ji} + e_j \bg_{ji})\big]  =  0 \ ,
\label{tadIj}
\eeq
where we have taken into account the constraints following from 
(\ref{jacqhpf}).

Unlike the situation with only NS, RR and geometric fluxes, we do not
have at the moment a prescription, e.g. by doing 
generalized dimensional reduction of a 
10-dimensional theory, to obtain the S-dual equivalent of the 
Bianchi identities. We have thus used as a guide the 
duality transformations. 
It would nevertheless be very interesting to have 
methods different from duality arguments in order to obtain the
complete set of constraints.

\section{\neq1  Minkowski minima with moduli fixed}
\label{secvacua}

In this section we intend to start analyzing the landscape of
vacua of the moduli potential when S-dual fluxes are
turned on. The essential new feature in the superpotential will 
be the presence  of terms (in type IIB langauge) of the form $ST_iP(U_j)$,
with $P(U_j)$ a cubic polynomial with integer flux coefficients.
The complete superpotential is the sum of 
eqs.~(\ref{wbtdual}) and (\ref{wpdetail}). 
To study generic supersymmetric and non-supersymmetric
vacua of the ensuing moduli potential is beyond the scope
of this work. As a first step towards exploring the effect
of S-dual fluxes we will only look here  for supersymmetric
Minkowski solutions in which the potential attains a minimum 
when the superpotential and its partial derivatives vanish. 

We will work in IIB/O3 and restrict to the case
with $T_i=T$, $U_i=U$ and fluxes given by (\ref{isofluxes}) 
and (\ref{giso}). The superpotential then simplifies to 
\beqa
\cw & = & e_0 + 3i e U - 3q U^2  +i m U^3 
\nonumber \\[0.2cm]
& + & S \big[ ih_0 - 3 a U + 3i \ba U^2  - \bh_0 U^3 \big]
\label{wisosdual} \\[0.2cm]
& + & 3T \big[ -ih - (2b+\b) U + i (2\bb + \bbet) U^2  + \bh U^3 \big]
\nonumber \\[0.2cm]
& + & 3ST \big[ -f +i (2g+\g) U +  (2\bg + \bgam) U^2  -i \ff U^3 \big]  \ .
\nonumber
\eeqa
In this isotropic case the tadpole cancellation conditions become
\beqa
N_{\rm D3} + \oh \big[ m h_0 - e_0 \bh_0 + 3qa + 3e \ba \big]  & = &  16 \ ,
\nonumber \\[0.2cm]
-N_{\rm D7} +  \oh \big[m h - e_0 \bh -
q (2b + \b)  -  e(2\bb + \bbet) \big]  & = &   0 \ ,
\label{isotad}  \\[0.2cm]  
-N_{\rm NS7} +  \oh \big[
h_0\ff  - \bh_0 f - \ba (2g + \g) + a(2\bg + \bgam) \big]& = & 0 \ .
\nonumber \\[0.2cm]
N_{\rm I7} +  \oh \big[e_0 \ff - mf +
q (2g + \g)  +  e(2\bg + \bgam) \big]  & = &   0 \ .
\nonumber
\eeqa
The fluxes must further satisfy the constraints derived in
sections \ref{ssecbianching} and \ref{ssecbianchisdual}.

It is helpful to make a change of variables
\beq
U= -i \rho \quad ; \quad S= -i \sigma \quad ; \quad 
T= -i \tau \ .
\label{nust}
\eeq
The superpotential (\ref{wisosdual}) then becomes
\beq
\cw = E_1 + \sigma E_2 + \tau E_3 + \sigma\tau E_4  \  ,
\label{genw}
\eeq
where the $E_i$ are cubic polynomials in $\rho$ given by
\beqa
E_1 & = & e_0 + 3 e \rho + 3q \rho^2  - m \rho^3 \ ,
\nonumber \\[0.2cm]
E_2 & = & h_0 + 3 a \rho - 3\ba \rho^2  - \bh_0 \rho^3  \ ,
\label{nwiso} \\[0.2cm]
E_3 & = & 3\big[ -h + (2b+\b) \rho - (2\bb + \bbet) \rho^2  + \bh \rho^3 \big]  \ ,
\nonumber \\[0.2cm]
E_4 & = & 3\big[ f - (2g+\g) \rho +  (2\bg + \bgam) \rho^2  -\ff \rho^3 \big]  \ .
\nonumber
\eeqa
The advantage is that now all coefficients are real, in fact integers.

The problem is to find solutions of
\beq
\cw = \frac{\p\cw}{\p\r}=\frac{\p\cw}{\p\sigma}=
\frac{\p\cw}{\p\tau} = 0 \ .
\label{minkon}
\eeq
To begin, let us review the known situation in which there are neither
$\cq$ nor $\cp$ fluxes \cite{kst}. In this case $E_3=E_4=0$ and $\cw$ does not
depend on $\tau$. From
$\p\cw/\p\sigma=0$ and $\cw=0$ we find $E_1=E_2=0$. The remaining equation
$\p\cw/\p\r=0$ gives $\sigma=-E_1^\prime /E_2^\prime$ (where prime denotes
derivative with respect to $\rho$). The task is to determine whether
$E_1=0$ and $E_2=0$ have a common root $\r=\r_0$ with $\r_0$
necessarily complex (so that $\re U \not=0$ at the minimum).
Now, $\r_0^*$ must also be a root because the $E_i$ have real coefficients.
Thus, the $E_i$ must factorize as \cite{kst}
\beq
E_i = (\r - \r_0) (\r - \r_0^*) (\mu_i \r + \nu_i)
\label{fpoli}
\eeq
with $\mu_i$ and $\nu_i$ some coefficients that depend on the fluxes. In \cite{kst}
it is shown that there are fluxes that allow such a factorization and
moreover lead to $\re S > 0$. These fluxes contribute to the $C_4$ tadpole
as D3-branes. 

In a similar spirit we can consider the case $\ov{\ch}_3=0$ and $\cp=0$
in which $E_2=E_4=0$ and $\cw$ does not depend on $\sigma$.
Now $E_1$ and $E_3$ must factorize as in (\ref{fpoli}). For $E_1$
this poses no problem because the fluxes $e_0, e, q$ and $ m$ are unconstrained
when $\cp=0$. However, for the coefficients of $E_3$ we have the Bianchi
conditions (\ref{hfz}). These can be satisfied taking for example
$h=0$, $\bb=0$ and $\b=-b$. To simplify we also take $\bbet=0$. Then,
$\r_0=i\sqrt{b/\bh}$ which requires $\bh b > 0$. Imposing that $E_1$
has the same root gives the conditions $e_0 \bh = 3 qb$ and $me_0=-9eq$
which can be satisfied with $m=e=0$.
It is also instructive to compute $\tau=-E_1^\prime/E_3^\prime$. We find
$\tau=q \r_0/b$. Then $\re T > 0$ requires $qb > 0$ and this implies that
the  $C_8$ tadpole is negative (same sign as for a D7-brane).  

To continue the systematic analysis we can set only the
$\cp$-fluxes to zero so that
$E_4=0$. Now minimization with $\cw=0$ requires $E_1=E_2=E_3=0$.
Moreover, the coefficients of $E_2$ and $E_3$ must verify Bianchi constraints.
Fulfilling the constraints by using the solution (\ref{bpnz}) leads
to $E_3$ without complex roots. Taking instead the constraint
solution (\ref{nadahqz}) we find that $E_3$ could have a complex
root only if $h_0 \ba > 0$, but it cannot be a simultaneous
root of $E_2$. 
The interesting conclusion is that to
fix all moduli in a supersymmetric Minkowski minimum,
within our class of solutions of Bianchi constraints, we have to go
beyond metric and non-geometric fluxes. 

We now come to the generic situation with all fluxes turned on. From 
$\p\cw/\p\sigma=0$ and $\p\cw/\p\tau=0$ we find
\beq
\tau=- \frac{E_2}{E_4} \quad ; \quad   \sigma=- \frac{E_3}{E_4}  \  .
\label{stsols}
\eeq
Substituting in $\cw=0$ and $\p\cw/\p\r=0$ then gives
\beq
E= E_1 E_4 - E_2 E_3 =0 \quad ; \quad E^\prime=0  \  .
\label{Pdefcond}
\eeq
Thus, this $E$ must have a double root $\r_0$, necessarily
complex. We also know that $E$ has real coefficients and 
is generically of order six in $\rho$. Hence, it can be
written as
\beq
E = 3(\r - \r_0)^2 (\r - \r_0^*)^2 (\a \r^2 + \delta \r + \ep) \ ,
\label{ppoli}
\eeq 
where $\a$, $\delta$, $\ep$ and $\r_0$ depend on the fluxes. 

To proceed further we will implement specific solutions for 
the Bianchi identities derived in section \ref{ssecbianchisdual}.
We consider different cases according to how we solve the
constraints on $\cp$ and $\cq$ fluxes alone. These cases are shown
in table \ref{solus}, where we have also displayed the solution 
to the constraint involving the $\ov{\cf}_3$ and $\ov{\ch}_3$
fluxes. 

\begin{table}[htb] \footnotesize
\renewcommand{\arraystretch}{1.25}
\begin{center}
\begin{tabular}{|c|c|c|c|c|}
\hline
Case &  $\cp \cdot \cp=0$  &  $\cq \cdot \cq=0$ &  $\cq \cdot \cp + \cp \cdot \cq=0$ &
$\cq \cdot \ov{\ch}_3 - \cp \cdot \ov{\cf}_3=0$ \\
\hline
 &   & $\b=-b$, $\bbet=-\bb$  &  &  $h_0 \bb=e\g + \ba h$ \\
\raisebox{2.5ex}[0cm][0cm]{1} &
\raisebox{2.5ex}[0cm][0cm]{$f=\ff=g=\bg=0$} & 
$h\bh=b \bb$ &
\raisebox{2.5ex}[0cm][0cm]{$b\g = h \bgam$} &  $\bh_0 h=q\g + a \bb$ \\
\hline
&  $\g=-g$, $\bgam=-\bg$  &  &  &   $mg= e\ff - \ba \bbet$ \\
\raisebox{2.5ex}[0cm][0cm]{2} &
$f\ff=g \bg$ & 
\raisebox{2.5ex}[0cm][0cm]{$h=\bh=b=\bb=0$} &
\raisebox{2.5ex}[0cm][0cm]{$f\bbet = g \b$} &  $q g= a \bbet - e_0 \ff$ \\
\hline
\end{tabular}
\end{center}
\caption{\small Solutions to Bianchi identities}
\label{solus}
\end{table}

Within our class of solutions the polynomials $E_3$ and $E_4$
always take a simpler form. For example, in case 1 they
are given by
\beq
E_3  =  \frac{3}{h}(\bb\r^2 + h)(b \r - h)
\quad ; \quad
E_4  =  \frac{3\g \r}{h}(b \r - h)  \ .
\label{e34}
\eeq
In each case we compute $E$ and check if it can be factorized in the form
(\ref{ppoli}). When this is possible we can determine $\r_0$.
Moreover, typically there will be relations among the fluxes.
We have limited ourselves to finding some solutions.
Details are presented below.

\bigskip
\noindent
{\bf Case 1}

\noindent
We find a class of minima with fluxes
satisfying the relations
\beq
q = 0 \quad ; \quad  
e_0 \g = 4 a h
\quad ; \quad 
m h_0 h \g = (e\g + 4 h \ba )(e\g + h \ba ) \  .
\label{frel1}
\eeq
Besides, $q=0$ implies $h \bh_0=a\bb$, with $\bb$ given in table \ref{solus}.
As free parameters we can then take $a$, $\ba$, $b$, $h_0$, $h$, $\g$ and $e$. They must
be such that the remaining dependent fluxes come out integers as well.
Furthermore, there are sign relations required for consistency. For example,
we find
\beq
\rho_0 = i |\r_0| \quad ; \quad |\r_0|^2=-\frac{h_0 h}{e\g + h \ba} \  .
\label{rsol1}
\eeq
This needs $h_0 h (e\g + h \ba) < 0$, then $U=\sqrt{-h_0 h/(e\g + h \ba)}$. 

The remaining moduli turn out to be
\beqa
S & = & \frac{2h}{\g U}  \ ,
\nonumber \\[0.2cm]
T & = &  \frac{h\big[h_0(2\ba h - e \g) - 2i a U(e\g + \ba h)\big]}{3\g U (e\g + \ba h) (h - ibU)} \ .
\label{stcase1} 
\eeqa
To guarantee $\re S > 0$ and $\re T > 0$ we need 
\beq
h \g > 0  \quad  ; \quad (2a b - 2 \ba h + e\g) > 0  \ .
\label{poscon1}
\eeq
For example, choosing
\beq
a=-8 \ ; \
b=-4 \ ; \
h=-4  \ ; \ \g=-4 \ ; \
\ba=12 \ ; \
h_0=8   \ ; \ e=-16   \ ,
\label{npsol1} 
\eeq
we find that all dependent fluxes are also even integers.
The moduli are determined to be
\beq
U=S=\sqrt{2} \quad ; \quad T=\frac19(14 \sqrt{2} + 16i) \ .
\label{modsol1} 
\eeq
It is also interesting to compute the tadpoles. In particular we find
\beq
N_{\D 3} -288 = 16 \ \ ; \ \ -N_{\D 7}+80 = 0 \ \ ; \ \
-N_{\rm NS7} + 40 = 0 \ \ ; \ \ N_{\rm I7} + 32 = 0 \ .
\label{tadsol1} 
\eeq
We observe the peculiar result that fluxes contribute to the
$C_4$ tadpole as O3-planes instead of D3-branes. 
However, this is not  generic. In other examples the $C_4$
flux tadpole comes out positive. 
The 8-form tadpoles can
have either sign, or even cancel, depending on the parameters.

Within the above class of vacua we can set $e=0$, implying
$h_0 \bb = \ba h$. The I7 tadpole cancels since now $q=e=0$. 
On the other hand, from the condition $\re S > 0$ we conclude 
that in this case  
fluxes always contribute to the $C_4$ tadpole as D3-branes.
Using also the condition for $\re T > 0$ shows that the flux
piece in the $C_8$ tadpole is positive (opposite sign as
D7-branes). The flux piece in the NS7 ($\wt{C}_8$) tadpole
is also positive. In fact, the flux 
tadpoles of $C_8$ and $\wt{C}_8$ are both positive and proportional
to $\re T$.
To give a numerical example, we can take $\ba=-h_0=2$, $U=1$, and 
\beq
a=-8 \quad ; \quad
b=-2 \quad ; \quad
h=-2  \quad ; \quad \g=-2 \quad ; \quad
S=2 \quad ; \quad T=\frac53-i \ .
\label{psol1} 
\eeq
It is easy to verify that the dependent fluxes are all even integers.

If we relax the condition $b\g = h \bgam$ we can find 
vacua with all 7-brane tadpoles zero but $\re T \not=0$.
To cancel these tadpoles from the beginning we impose 
$e=q=0$, together with $mh=e_0 \bh$ and $a\bgam = \ba \g$.  
We then find a solution 
provided that we also fulfill the relations
\beq
\frac{h a}{b \ba} > 0 \quad ; \quad
a^2 = - h_0 \ba  \quad ; \quad
m\g = -\frac{\ba (a h - b h_0)^2}{h h_0^2}  \  .
\label{zerod7ns7extra}
\eeq
As independent fluxes we can now choose $\ba$,  
$b$, $a$, $h$ and $\g$. 
For the moduli we obtain
\beqa
U & = & \sqrt{\frac{h a}{b \ba}} \ ,
\nonumber \\[0.2cm]
S & = & \frac{-ib (a + i\ba U)^2}{\g a^2}   \ ,
\label{stcase1tc} \\[0.2cm]
T & = &  \frac{(a - i \ba U)^2}{3\g \ba U}   \ .
\nonumber 
\eeqa
To have $\re S > 0$ it suffices to impose $h \g > 0$. The
flux contribution to the D3 tadpole is then positive.
We can also arrange to have all fluxes to be integers
while $\re T$ is positive and large.
However, the flux D3 tadpole will also be large.

\bigskip
\noindent
{\bf Case 2}

\noindent
There is a type of solutions with
free parameters  $\b$, $\bbet$, $e_0$, $e$, $\ff$, $g$ and $a$, with remaining fluxes
determined by
\beq
\ba = 0 \quad ; \quad h_0 \bbet = 4 e g
\quad ; \quad \bh_0 \bbet = \frac{\ff(4 e_0 \ff - 3 a\bbet)}{g} \  .
\label{frel3}
\eeq
Notice that then $m=e \ff/g$ and $q$ is given in table \ref{solus}.
There is a sign condition $g \ff < 0$, then $U=\sqrt{-g/\ff}$. 
We further obtain
\beqa
S & = & \frac{\bbet}{2\ff U}  \ ,
\nonumber \\[0.2cm]
T & = &  \frac{U}{3g}   \frac{\big[2 e \ff U + i (2 e_0 \ff - 3 a\bbet)\big]}{(\bbet U + i \b)} \ .
\label{stcase3} 
\nonumber
\eeqa
It is easy to check that $\re S > 0$ and $\re T > 0$ are positive as long as
\beq
g \bbet < 0  \quad ; \quad \left(2e g + 3a \b - \frac{2e_0 \b\ff}{\bbet} \right) > 0  \ .
\label{poscon3}
\eeq
For an illustrative example, consider the parameters
\beq
\ff=2 \ ; \
\b=8 \ ; \
\bbet=8  \ ; \ g=-2 \ ; \
a=28 \ ; \
e_0=16   \ ; \ e=96   \ .
\label{npsol2} 
\eeq
The moduli are then fixed as
\beq
U=1 \quad ; \quad S=2 \quad ; \quad T=\frac13(7 + 31i) \ .
\label{modsol2} 
\eeq
For the flux-induced tadpoles we obtain
\beq
N_{\D 3}+ 32 = 16 \quad ; \quad  N_{\rm I7} + 112 = 0 \ .
\label{tadsol2} 
\eeq
The flux contribution to $C_8$ and $\wt{C}_8$ tadpoles is zero.

The I7 tadpoles cancel when $a=0$. Then it is simpler to show
that the free parameters can be chosen so that all other fluxes are integers
while $\re S$ and $\re T$ are large and positive. 
The sign relations among the parameters imply that the $C_4$, $C_8$ and 
$\wt{C}_8$ flux tadpoles are positive. The latter two
are proportional to $\re T$.

Similar to case 1, if we relax the condition $f\bbet = g \b$ we can find 
minima with all 7-brane tadpoles zero but $\re T \not=0$.

Other solutions of the Bianchi identities can be obtained
by combining the building blocks of table \ref{solus}.
For example, $\cp\! \cdot \!\cp =0$ can be fulfilled as
in case 2, and  $\cq\! \cdot \!\cq =0$ as in case 1. Then
the solution of $\cq\! \cdot \!\cp + \cp \!\cdot \! \cq =0$ can
be written as $(bf-gh)(h\ff - g\bb)=0$. Now, if
$h \ff =g \bb$, $E_3$ and $E_4$ have a common
quadratic factor and it can then be  
shown that the polynomial $E$ cannot be 
factorized as needed. 
When $bf=gh$, to avoid $\re S=0$ it must be that $U$ is
necessarily complex. In this more complicated
case we have not been able to find supersymmetric
Minkowski minima.

In summary, some differences compared to the
type IIB results in ref.\cite{kst} 
without non-geometric nor S-dual fluxes are 
evident. The situation now is rather more involved but
still we have found some concrete results.
For simplicity we have analyzed the case with isotropic
fluxes and moduli $T_i=T$, $U_i=U$.
We find Minkowski \neq1 vacua in which not only the dilaton 
and  complex structure fields are fixed but also
the  K\"ahler modulus $T$ is fixed.
However, if we  analyze the more general case with 
independent $T_i$, $U_i$ fields, generically 
only one linear combination of the K\"ahler moduli 
$T_i$ is fixed. This is due to the fact that the 
superpotential is only linear in the $T_i$ and
essentially only depends on a linear combination 
of these moduli. 

When S-dual backgrounds are switched on, 
the contribution from fluxes to the tadpole of the RR
$C_4$ form can have either sign depending on
the flux values. This is a surprising result.
We know that in absence of S-dual fluxes the $C_4$ tadpole due
to $\ov{\ch}_3$ and $ \ov{\cf}_3$ fluxes 
consistent with the imaginary self-dual condition
needed for supersymmetry is always positive \cite{gkp, kst}. 
Concerning the $C_8$ tadpole, if only
non-geometric fluxes are present, as in a toy example
with $\cw$ depending only on $U$ and $T$, the flux tadpole
is negative (same sign as D7-branes). 
However, in presence
of S-dual backgrounds the flux contribution can be positive 
(same sign as O7-planes), negative, or even vanish. 
An analogous result occurs in AdS type IIA  vacua
with metric fluxes in \cite{cfi}.
The fact that fluxes may contribute to tadpoles as
orientifold planes may be useful for model-building
as already emphasized in \cite{cfi}.

The value of the real parts of the dilaton $S$ and the size modulus $T$ 
may be made large by appropriately choosing the fluxes.
This is in general required to maintain perturbative
values for the couplings and the validity of the supergravity
approximation. On the other hand, 
in our supersymmetric Minkowski vacua
$\re S$ and $\re T$  cannot be made arbitrarily large
because in general they are tied to RR tadpoles induced by the fluxes.
This could be avoided in AdS vacua, as occurs in the IIA/O6 models
without \cite{DeWolfe} and with metric fluxes \cite{cfi}.
We have not found non-trivial supersymmetric Minkowski vacua
with full cancellation of RR tadpoles, although in some cases 
one can find flux combinations with vanishing contribution 
to some tadpoles. 
We assume that localized sources of different kinds 
may be added to the theory rendering it  tadpole free, as 
it happened in the simpler examples in ref~\cite{cfi}.
In this connection, notice that if we want to add D3 and/or
D7-branes to vacua like these, 
the existence of undetermined K\"ahler $T_i$ moduli 
may in fact be necessary, as emphasized elsewhere \cite{cfi}. 
In particular, in the worldvolume of generic branes 
(not on top of orientifold planes) live $U(1)$ groups that 
may have triangle anomalies in four dimensions. The $U(1)$
gauge vectors become massive
through a generalized Green-Schwarz mechanism by swallowing 
some linear combination of the $T_i$. Thus, for this mechanism to work,  
some $T_i$ fields should be left unfixed by the fluxes.
This turns out to be related
to the requirement of absence of Freed-Witten brane
worldvolume anomalies \cite{cfi}.

\section{Generalized Duality Invariant Superpotentials}
\label{secgeneral}

In previous sections we have described fluxes present in different 
T-dual type II  orientifolds. The closed string sector
of all these theories, before the
addition of fluxes, gives rise at low-energies to an effective 
\neq4 supergravity theory (or \neq1 if we further perform
a $\Z_2\times \Z_2$ twist). We would like to
compare now to the results obtained from other string constructions
having an analogous  low-energy structure.
Specifically,  we would like to compare to the flux-induced superpotential
in analogous heterotic compactifications as well as in  certain 
compactifications of M-theory on simple twisted 7-tori \cite{ap}.

\subsection{M-theory on a twisted 7-tori}
\label{ssecmth}

We consider here the $G_2$-holonomy manifolds $X_7$ obtained 
as certain $\Z_2\times \Z_2\times \Z_2$ orbifolds of the 7-torus,
$X_7=\T^7/\Z_2\times \Z_2\times \Z_2$  \cite{Joyce}. We will follow  the
results and notation used in ref.\cite{ap}. One has seven complex 
moduli fields $M_I(x)$, $I=1,...,7$. They may be defined in terms of 
the complexified $G_2$-form
\begin{equation}
C_3\ +\ i \, {\bf \Phi}_3\ = \ i M_I(x)\ \phi^I(y)
\end{equation} 
where $\phi^I\in H^3(X_7)$, $C_3$ is the M-theory 3-form 
and ${\bf \Phi}_3=\re M_I(x)\phi^I(y)$, with 
$\re M_I(x)$  parameterizing the volume of the 7
invariant 3-cycles in $X_7=T^7/\Z_2\times \Z_2\times \Z_2$.
We will now consider the addition of metric fluxes in this
toroidal model. This is a Scherk-Schwarz reduction which proceeds
in a way analogous to that described for type IIA 
orientifold compactifications. In particular we replace the 
differentials $dy^P$, $P=1, \cdots, 7$,  by twisted forms $\eta^P$ satisfying
\begin{equation}
d\eta^P = -\oh \omega^P_{MN} \eta^M \wedge \eta^N \ ,
 \omega^P_{[MN} \omega^S_{R]P} = 0 \  \ 
\label{jacm}
\end{equation}
where one also has  $\omega^P_{PN}=0$ \cite{ss, km}.
Among these metric fluxes $\omega^P_{MN}$,  only twenty-one are invariant under
the twists.  In addition we consider the presence of seven 4-form 
backgrounds $g_{IJKL}$ corresponding to fluxes of the M-theory 3-form.
The presence of these two types of fluxes gives rise to a superpotential
\cite{bw,MT,ap} 
\begin{equation}
W_{7}\ =\ \frac {1}{4} \int_{X_7} (C+i{\bf \Phi })\wedge [g+\frac {1}{2}
d(C+i{\bf \Phi })] \ +\ \frac {1}{4} \int_{X_7} G_7
\label{supermth}
\end{equation}
Here $G_7$ is the flux of the 3-form dual.  Expanding this 
superpotential in terms of the seven moduli in type IIA notation
\cite{ap} one obtains:
\beqa
W_7\ & = &\ g_{567891011} \ + \ i
(g_{78910}T_1+g_{56910}T_2+g_{5678}T_3) 
\ +\  \label{supermiia} \\ 
&+ & i(g_{57911}S-g_{581011}U_1-g_{671011}U_2-g_{68911}U_3) \nonumber \\
&+ & (\omega_{910}^{11}T_1T_2+\omega_{56}^{11}T_2T_3+\omega_{78}^{11}T_1T_3)
-S(\omega_{79}^6T_1+\omega_{95}^8T_2+\omega_{57}^{10}T_3) \nonumber \\
&+& (\omega_{810}^6T_1U_1+\omega_{106}^8T_2U_2+\omega_{68}^{10}T_3U_3)
\ -\ (\omega_{710}^5T_1U_2+\omega_{89}^5T_1U_3+\omega_{105}^7T_2U_1) 
\nonumber \\
&+& (\omega_{96}^7T_2U_3+\omega_{58}^9T_3U_1+\omega_{67}^9T_3U_2) \nonumber \\
&-& S(\omega_{511}^6U_1+\omega_{711}^8U_2+\omega_{911}^{10}U_3)
\ +\ \omega_{1011}^9U_1U_2+\omega_{611}^5U_2U_3+\omega_{811}^7U_1U_3
\nonumber
\eeqa
All terms in this superpotential, except for those in the last line,
may be understood in terms of ordinary RR and NS backgrounds 
in the type IIA orientifold supplemented by metric fluxes.
Indeed, all those terms correspond to the fluxes $e_0$, $e_i$, $h_0$,
$h_i$, $q_i$, $a_i$, and $b_{ij}$, described in chapter 2. The absence of a 
$T_1 T_2 T_3$ term (type IIA mass parameter $m$) is expected since in the 
M-theory scheme considered massive IIA supergravity  does not
arise. 

The new terms appearing in the last line are interesting.
The first three  correspond to the S-dual fluxes $f_i$ introduced
before in order to maintain S-duality in the IIB orientifold version
of this model. Thus one has the interesting result that the $f_i$ fluxes
introduced before may be understood as certain ordinary metric fluxes
\beq
f_i \ =\ \omega_{K11}^{K+1} \ \ \  ;\ \ K=5,7,9 
\label{quesf}
\eeq
in an M-theory version of the same model.  
On the other hand the last three terms, bilinear in the $U_i$ ($T_i$)
in the IIA (IIB) version,  
are new and are absent even in the extended set of flux-induced
superpotential terms discussed in previous chapters.
This suggests that there is an even bigger set of flux degrees of freedom 
to be considered. We will see now that the presence of new terms
bilinear and cubic in the $U_i$'s are also expected if we
consider fluxes in the heterotic version of the same
class of models.

\subsection{Heterotic fluxes}
\label{ssechet}

The type IIA orientifold with O6-planes is T-dual to the 
type IIB orientifold with O9-planes, i.e. type I string theory.
On the other hand we know that type I is related by S-duality
to the $SO(32)$ heterotic string. Therefore,  it is interesting to compare 
the induced superpotentials in both theories. 
Flux-induced heterotic superpotentials have been analyzed in
\cite{Becker1, Becker2, Cardoso, Gurrieri}. 
It has been argued that heterotic H-flux forces the internal
manifold $X_6$ to be non-K\"ahler with $dJ \not=0$. 
Both effects produce a superpotential
\beq
W_{\rm het}\ =\ \int_{X_6} \Omega \wedge (\ov{H}_{\rm het}\ +\  dJ_c)
\label{superhet}
\eeq
An example of non-K\"ahler manifold is the twisted torus in
which $dJ_c = \om J_c$, where $\om$ are the metric fluxes.

It is interesting to evaluate $W_{\rm het}$ in the case of
compactification on a factorized $\T^6$ with arbitrary
metric fluxes on top. The $H$-flux is a generic 3-form, namely
\beq
\ov{H}_{\rm het}  =  -e_0 \a_0 + m\b_0 - \sum_{i=1}^3 (q_i\a_i + e_i \b_i) \ .
\label{hhet} 
\eeq
Our choice of parameters is dictated by the fact that by S-duality
$\ov{H}_{\rm het}$ is equal to the RR flux, given in table \ref{rriib},
of IIB with O9-planes, alias type I. Moreover, the heterotic metric fluxes
are the same as those as in IIB/O9 shown in table \ref{nongeoiib}. 
This means $\om_{\rm het}=\bom$. 
We also need to use that in the toroidal compactification
the heterotic complex structure moduli coincide with the geometric
parameters, i.e. $U_i = \tau_i$. The K\"ahler moduli arise from
$J_c=i\sum_j T_j \om_j$. Putting all pieces together we find
\beqa
W_{\rm het}  & = & m + i\sum_{i=1}^3 q_i U_i
+ e_1 U_2 U_3 + e_2 U_1 U_3 + e_3 U_1 U_2 - i e_0 U_1 U_2 U_3
\label{whet}  \\[0.2cm]
& +  & \sum_{i=1}^3 T_i \big[ -i\bh_i  + \sum_{j=1}^3
\bb_{ji} U_j  + i b_{1i}  U_2 U_3  + i b_{2i} U_1 U_3 
+ i b_{3i}  U_1 U_2 -  h_i U_1 U_2 U_3  \big]   \ .
\nonumber 
\eeqa
Superpotentials of this kind have been recently considered in \cite{deCarlos}.
With isotropic choice of fluxes $W_{\rm het}$ agrees with results of \cite{dkpz}. 

Comparing with (\ref{wo9tdual}) shows that $W_{\rm het}$ matches $W_{\rm O9}$ 
except for the terms linear in $S$ that are due to non-geometric fluxes 
$\bR$ in IIB/O9.  
Additional $S$-dependent terms in $W_{\rm O9}$ will appear if 
S-dual fluxes are included (T-dual to the $\cp$).
Thus, we conjecture that analogous dilaton-dependent superpotential
terms will emerge in the heterotic side from new flux
degrees of freedom $R_{\rm het}$ and $P_{\rm het}$.

\subsection{Fluxes and $SL(2,\Z)^7$ duality invariance}

We have just argued that dilaton-dependent terms in $W_{\rm het}$ 
would arise from heterotic fluxes $R_{\rm het}$ and $P_{\rm het}$. Now,
there are reasons to believe that this is not the whole story. 
In particular, we know that 
4-dimensional compactified heterotic strings are self-T-duality
invariant. As a consequence, the complete K\"ahler function,
$\cg=K+ \log|\cw|^2$, should be invariant
under the $SL(2,\Z)^3$ heterotic T-duality symmetries
\cite{flst}.  
In order to be so, the superpotential $W_{\rm het}$ should
transform appropriately. It is easy to convince oneself that
this demands terms quadratic and
cubic in the K\"ahler moduli $T_i$ in $W_{\rm het}$
\footnote{Presence of such terms has also been recently pointed out in
\cite{az}.}.
We already observed in section \ref{ssecmth} that such 
new quadratic terms seem to be present. We now see that
they are also required to help in restoring heterotic self-T-duality.

\begin{table}[htb] \footnotesize
\begin{center}
\begin{tabular}{|c|c|c|c|c|}
\hline
{} &  \multicolumn{4}{c|}{components} \\
\raisebox{2.5ex}[0cm][0cm]{Flux} & \multicolumn{4}{c|}{induced terms} \\
\hline
{}& $m$ & $q_i$ & $e_i$ & $e_0$ \\
\raisebox{2.5ex}[0cm][0cm]{$\ov{\cf}_3$}  & $U^3$ & $U^2$ & $U$ & $1$ \\
\hline 
{}& $\bh_0$ & $\ba_i$ & $a_i$ & $h_0$ \\
\raisebox{2.5ex}[0cm][0cm]{$\ov{\ch}_3$}  & $SU^3$ & $SU^2$ & $SU$ & $S$ \\
\hline 
{}& $\bh_i$ & $\bb_{ij}$ & $b_{ij}$ & $h_i$ \\
\raisebox{2.5ex}[0cm][0cm]{$\cq$}  & $TU^3$ & $TU^2$ & $TU$ & $T$ \\
\hline 
{}& $\ff_i$ & $\bg_{ij}$ & $g_{ij}$ & $f_i$ \\
\raisebox{2.5ex}[0cm][0cm]{$\cp$}  & $STU^3$ & $STU^2$ & $STU$ & $ST$ \\
\hline
\hline
{}& $m^\prime$ & $q^\prime_i$ & $e^\prime_i$ & $e^\prime_0$ \\
\raisebox{2.5ex}[0cm][0cm]{$\ov{\cf}_3^\prime$}  & $T^3U^3$ & $T^3U^2$ & $T^3U$ & $T^3$ \\
\hline 
{}& $\bh^\prime_0$ & $\ba^\prime_i$ & $a^\prime_i$ & $h^\prime_0$ \\
\raisebox{2.5ex}[0cm][0cm]{$\ov{\ch}_3^\prime$}  & $ST^3U^3$ & $ST^3U^2$ & $ST^3U$ & $ST^3$ \\
\hline 
{}& $\bh^\prime_i$ & $\bbp_{ij}$ & $b^\prime_{ij}$ & $h^\prime_i$ \\
\raisebox{2.5ex}[0cm][0cm]{$\cq^\prime$}  & $T^2U^3$ & $T^2U^2$ & $T^2U$ & $T^2$ \\
\hline 
{}& $\ff^\prime_i$ & $\bgp_{ij}$ & $g^\prime_{ij}$ & $f^\prime_i$ \\
\raisebox{2.5ex}[0cm][0cm]{$\cp^\prime$}  & $ST^2U^3$ & $ST^2U^2$ & $ST^2U$ & $ST^2$ \\
\hline
\end{tabular}
\end{center}
\caption{\small IIB/O3 fluxes and their induced terms}
\label{alliib}
\end{table}

To elaborate the point, it helps to look at table \ref{alliib}.
The upper half shows the fluxes that we have already encountered in IIB/O3,
together with the characteristic term that they induce.
$W_{\rm het}$ contains the monomials due to fluxes of type 
$\ov{\cf}_3$ and $\cq$ in the table (upon $h_i \leftrightarrow \bh_i$, 
$e_0  \leftrightarrow m$, etc.). In order to realize heterotic
self T-duality new fluxes of type $\ov{\cf}_3^\prime$ and $\cq^\prime$
need to be added.  
Note that terms quadratic in $T_i$'s, already manifest in the
M-theory analysis, come from the flux $h^\prime_i$.
Similarly, self-T-duality of $S$-dependent terms, due
to $\ov{\ch}_3$ and $\cp$ in IIB/O3, requires new fluxes 
$\ov{\ch}_3^\prime$ and $\cp^\prime$. 

All these duality connections among fluxes in different
dual incarnations of the same theory suggest that the 
complete underlying theory is invariant under 
the $SL(2,\Z)^7$ transformations corresponding to the
seven untwisted moduli in this model. The general flux superpotential
will then be a polynomial of degree up to seven 
on the moduli $M_I=(S,T_1,T_2,T_3,U_1,U_2,U_3)$ and at
most linear on any of them. One can write this superpotential in the
form:
\beq
W_{Flux}\ =\ 
\sum_{n=0}^7\ D^{(n)}_{i_1...i_n}\ M_{i_1}...M_{i_n}
\label{cojopot}
\eeq
where the $D^{(n)}$ are integer coefficients associated to
generalized fluxes\footnote{General superpotentials of this type 
were considered previously in \cite{dkpz}
from the point of view of gauged \neq4 supergravity.}.
Under $SL(2,\Z)_X$ the modulus $M_X$ transforms as
\beq
M_X \to \frac {(k_X M_X-i\ell_X M_X)}{(im_X M_X + n_X)}
\quad ; \quad k_X n_X -\ell_X m_X=1
\quad ; \quad k_X,\, \ell_X, \, m_X, \, n_X \in \Z  \ .
\label{modular}
\eeq
The toroidal K\"ahler potential transforms like
\beq 
K\ \to  K\ +\  \log|im_X M_X+ n_X|^2
\eeq
and the complete K\"ahler function is invariant   
as long as the fluxes $D^{(n)}$ transform like
\beq
(D^{(n)}_{ijk..}, D^{(n+1)}_{xijk..})
\longrightarrow (D^{(n)}_{ijk..}, D^{(n+1)}_{xijk..})
 \left(
\begin{array}{cc}
n_X & m_X \\
\ell_X & k_X   
\end{array}
\right)       \ \ .
\label{fluxtrans}
\eeq
The fluxes $D^{(n)}$ may be viewed as symmetric tensors
of $n$ indices, with all diagonal components vanishing,
thus with binomial coefficient ${7 \choose n}$ independent
components. 
Hence, the total number of generalized fluxes is
$\sum_{n=0}^7 {7 \choose n}=2^7=2^{(h_{21}+h_{11}+1)}$. 
They provide the 128 components of a 
representation $(2,2,2,2,2,2,2)$ under $SL(2,\Z)^7$. 
As explained in the appendix, this in turn may be 
embedded into the spinorial ${\bf {128}}$ of $SO(7,7;\Z)$.
One can decompose the two Weyl spinors of fluxes accordingly to its $SU(7)$
tensorial structure
\beqa
\mathbf{64}& = &\mathbf{1}~\oplus~\mathbf{7}~\oplus~\mathbf{21}~\oplus~\mathbf{35}
\nonumber \\
\mathbf{64}' & = &\mathbf{1}'~\oplus~\mathbf{7}'~\oplus~\mathbf{21}'~\oplus~\mathbf{35}'
\label{su7decomp}
\eeqa
The components of each representation are then given by
\begin{center}
\begin{tabular}{c|ccccccc}
Rep.&\multicolumn{7}{c}{Flux Components}\\ \hline
$\! \mathbf{1}$  &        &           &              &$e_0$    &            &            & \\
$\mathbf{7}'$ &        &           &$e_i$         &$h_0$    &$h_i$       &            & \\
$\! \! \mathbf{21}$ &        &$q_i$      &$a_i$         &$b_{ij}$ &$f_i$       &$h'_i$& \\
$\mathbf{35}'$  &$m$ & $\bar{a}_i$&$\bar{b}_{ij}$& $ g_{ij}$
&$b'_{ij}$& $f'_i$&$e'_{0}$\\
$\! \! \mathbf{35}$
&   $\bar{h}_0$&$\bar{h}_i$&$  \bar{g}_{ij}$ & 
$\bar{b}'_{ij}  $&$g'_{ij}  $ & $e'_{i}$  & $h'_0$ \\
$\mathbf{21}'$ &        &$\bar{f}_i$      &$\bar{h}'_i$         &$\bar{g}'_{ij}$ &$q'_i$       &$a'_i$& \\
$\! \mathbf{7}$ &        &           &$\bar{f}'_i$         &$m'$    &$\bar{a}'_i$       &            & \\  
$\mathbf{1}'$  &        &           &              &$\bar{h}'_0$    &            &            &     
\end{tabular}
\end{center} 
Note that in the  M-theory setting described above, only the representations
$\mathbf{1}$, $\mathbf{7}'$ and $\mathbf{21}$ appear explicitly \cite{ap}.

In terms of component fluxes the full duality covariant superpotential may be
written as
\beqa
W_{Flux} & = & e_0-i\sum_{i=1}^3 h_iT_i+\frac{1}{2}\sum_{l\aneq m\aneq
n} h'_lT_mT_n+i e'_0 T_1T_2T_3 \label{todo} \\
& + & \bigg(ih_0-\sum_{i=1}^3 f_iT_i-\frac{i}{2}\sum_{l\aneq m\aneq
n} f'_lT_mT_n - h'_0 T_1T_2T_3 \bigg)S \nonumber \\
& + & \sum_{i=1}^3 \left[\bigg(-a_i+i\sum_{j=1}^3g_{ij}T_j
-\frac{1}{2}\sum_{l\aneq m\aneq n} g'_{il}T_mT_n + ia'_i T_1T_2T_3 \bigg)S \right. \nonumber \\
& + & \left.  ie_i-\sum_{j=1}^3 b_{ij}T_j-\frac{i}{2}\sum_{l\aneq m\aneq n}b'_{il}T_mT_n
-e'_iT_1T_2T_3 \right] U_i \nonumber\\
& + &  \frac{1}{2}\sum_{r\aneq s\aneq t}\left[\bigg(i\bar{a}_r+\sum_{j=1}^3\bar{g}_{rj}T_j+\frac{i}{2}
\sum_{l\aneq m\aneq n} \bar{g}'_{rl}T_mT_n - \bar{a}'_rT_1T_2T_3 \bigg)S \right. \nonumber\\
& - &  \left. q_r+i\sum_{j=1}^3\bar{b}_{rj}T_j
-\frac{1}{2}\sum_{l\aneq m\aneq n}\bar{b}'_{rl}T_mT_n + iq'_r
T_1T_2T_3\right]U_sU_t \nonumber \\
& + & \left[-\bigg(\bar{h}_0+i\sum_{j=1}^3\bar{f}_{j}T_j-\frac{1}{2}\sum_{l\aneq
m\aneq n}\bar{f}'_lT_mT_n + i\bar{h}'_0T_1T_2T_3\bigg)S \right. \nonumber\\
& + &  \left. im+\sum_{j=1}^3\bar{h}_{j}T_j+\frac{i}{2}\sum_{l\aneq m\aneq
n}\bar{h}'_lT_mT_n- m'T_1T_2T_3\right]U_1U_2U_3
\nonumber
\eeqa
The complexity  of this superpotential makes its analysis difficult, except in particular 
cases like those we have discussed in previous chapters.
In any event it is clear that there are many parameters
which should allow for new possibilities in fixing moduli.
It is important to remark that these 128 flux degrees of freedom are not
independent. We already saw how Bianchi identities and RR tadpoles
strongly restrict the possible fluxes in the simpler case
with 64 degrees of freedom. In the most general case analogous  
constraints should be fulfilled. It would be interesting to have
close expressions for these constraints in the more
general case.

Note that the above discussion does not imply that the effective 
action has full $SL(2,\Z)^7$ duality invariance. Indeed,
generic fluxes   break these symmetries. 
Rather, the above discussion  shows how the presence of each particular flux 
explicitly breaks the duality symmetries.
As we have seen, some of these flux degrees of freedom have a
simple interpretation as metric fluxes or explicit RR or NS
backgrounds in some particular version (type IIA or IIB orientifolds, 
heterotic, M-theory orbifold, ...) of compactified string theory.
Some other fluxes do not admit a simple geometric interpretation
and yet others are implied by  type IIB S-duality and/or 
heterotic self-T-dualities. Yet all of the 128 fluxes 
may  in general be present in the complete  underlying theory.

\section{Final comments and conclussions}

One of the main purposes of this work has been to study
the duality properties of the  flux degrees of freedom
in type II  \deq4 orientifolds, 
 as well as  in other related string vacua.
The addition of non-geometric fluxes restores
T-duality between IIB and IIA theories but spoils
type IIB S-duality.  We have seen how including  
new S-dual degrees of freedom this symmetry may 
be recovered. Once that is done, extra moduli dependent
terms appear in the effective superpotentials.
Taking into account these new terms we
 were able to find type IIB
 Minkowski \neq1 vacua in which not only
dilaton and complex structure but also some K\"ahler
moduli are fixed. In these classes of vacua we find
that fluxes may contribute to RR tadpoles 
with the same or opposite sign to that of D3- and D7-branes,
depending on the flux choice. This fact was already
found for type IIA  AdS vacua in \cite{cfi} and may
be relevant also for model-building.
We leave a  more systematic analysis and search for
other minima for future work.

The new and old fluxes are subject to a number of 
Bianchi and RR tadpole cancellation conditions.
We made use of S-duality $SL(2,\Z)$
transformations to deduce the form of the
new conditions involving all these fluxes.
It would be clearly interesting to derive those conditions
from other arguments independent from dualities.
It would also be important to understand the structure
 of branes which may be added in these generalized backgrounds
and possible constraints which they may suffer.
It is known that fluxes may give rise to
anomalies in the world-volume of branes and similar effects
are expected in the presence of new generalized fluxes.

Dualities relating these theories to heterotic and M-theory
compactifications suggest the existence of yet further flux
degrees of freedom, giving rise to yet more
terms in the effective superpotential. In our toroidal
examples a fully $SL(2,\Z)^7$ covariant superpotential
implies  the existence of   $2^7$
fluxes. The general superpotential contains all
possible monomials of the seven moduli which are at most
linear in any of them, with integer coefficients given by the
$2^7$ fluxes.
Many points remain to be better understood.  
It would be important to examine  the origin and structure of the
novel  S-dual fluxes as well as ways to understand the generalized
constraints on fluxes. The same applies to the extra
 flux degrees of freedom which might be required to get consistency
with the  full underlying duality symmetries.

Although we have
concentrated on a particular class of toroidal orientifolds we
believe that many of the points discussed (like e.g. the explicit
expressions of superpotentials in terms of integrals of fluxes
over the compact space) should have a more general validity.
What we find seems to indicate the existence of a large number
of flux degrees of freedom ($2^{(1+h_{21}+h_{11})}$  in our
examples) giving rise to a very rich superpotential in which
most or perhaps all moduli might be fixed.

\vspace*{1cm}

{\bf \large Acknowledgments}

We thank E. Andr\'es, F. Marchesano, T. Ort\'{\i}n,  
S. Theisen, and especially A. Uranga for useful discussions.
G.A. thanks the Departamento de F\'{\i}sica Te\'orica, Universidad 
Aut\'onoma de Madrid, for hospitality.
G.A. work is partially supported by   ANPCyT grant PICT11064.
The work of P.G.C. is supported by  the  Ministerio de Educaci\'on y Ciencia (Spain) 
through a FPU grant. P.G.C. thanks the Perimeter Institute for hospitality while
working on this paper.
A.F. thanks the Max-Planck-Institut f\"ur Gravitationsphysik
for hospitality while preparing this work. 
This work has been partially supported by the European Commission under
the RTN European Program MRTN-CT-2004-503369 and the CICYT (Spain).

\newpage

\section*{A. Spinorial embedding of background fluxes. }
\label{appA}
\setcounter{equation}{0}
\renewcommand{\theequation}{A.\arabic{equation}}

We have seen in section \ref{secgeneral} how the generalized 
duality invariant superpotential presents a $SL(2,\Z)^7$ symmetry. 
In this appendix we describe in detail how the fluxes
are arranged into this structure and their embedding into the spinorial 
representation of $SO(7,7;\Z)$.

Each of the seven $SL(2,\Z)_X$ factors consists of two generators
\beq
{\cal S}_{X,1} =\left(
\begin{array}{cc}
1 & 1 \\
0 & 1   
\end{array}
\right) \quad ; \quad  {\cal S}_{X,2} = \left(
\begin{array}{cc}
0 & -1 \\
1 & 0   
\end{array}
\right)   \ 
\eeq
acting on the modulus $M_X$. From equation 
(\ref{modular}) one can see that ${\cal S}_{X,1}$ corresponds to 
shifts on the corresponding axion and ${\cal S}_{X,2}$ to M-duality 
$M_X \to 1/M_X$.

The set of fluxes, denoted ${\mathbb G}$, contains 
128 weights of the form $(\pm  ,\pm ,\pm  ,\pm ,\pm  ,\pm ,\pm )$,
where $\pm$ stands for $\pm \oh$.
The transformation $M_X \to 1/M_X$ is simply given by
\beq
{\cal S}_{X,2}(n_1,\ldots,n_X,\ldots,n_7)=\textrm{Sign}(n_X)(n_1,\ldots,-n_X,\ldots,n_7) \ .
\label{mdualflux}
\eeq
Thus, eq.(\ref{todo}) transforms in such a way that the full
supergravity scalar potential is invariant under the
${\cal S}_{X,2}$ generators.   
 The resulting map between weights and flux components is 
presented in table \ref{mapa}. 
We see, for instance,  that $\ov{\cf}_3$ fluxes (table \ref{rriib}) 
correspond to $(+,+,+,+,{\underline {\pm,\pm,\pm}})$  while  $\ov{\ch}_3 $ fluxes 
(table  \ref{nsiib}) are
  represented by $(-,+,+,+,{\underline {\pm,\pm,\pm}})$ spinorial weights.  

>From table \ref{mapa} we can easily read the action of the duality  
group in the different fluxes. 
Notice also that the duality transformations can be easily obtained by expressing the 
$SL(2,\Z)$ generators in terms of lowering and raising operators.
Namely, ${\cal S}_{X,2}={\cal S}_{X,+}-{\cal S}_{X,-}$ and  ${\cal S}_{X,1}={\cal I}+{\cal S}_{X,-}$. 
Thus, for instance, ${\cal S}_{1,2} \ov{\cf}_3 =-\ov{\ch}_3$, corresponds to the  
S-duality transformation (\ref{sl2zflux}).

It is interesting to note how half of the 
degrees of freedom of each of the two Weyl spinors on which ${\mathbb G} = 
{\mathbf {64}}\oplus {\mathbf {64}}'$ can be decomposed correspond to RR fluxes, 
whereas the other half are generalized NS fluxes. Of these, half are heterotic 
and half are ordinary fluxes, thus giving a very symmetric structure.

\begin{table}[!ht]
\begin{center}
\begin{tabular}{c|c||c|c}
Flux parameter & Weight & Flux parameter & Weight\\
\hline \hline $\bar{h}'_0$ & $(-,-,-,-,-,-,-)$& $e_0$ & $(+,+,+,+,+,+,+)$\\
$h_0$ & $(-,+,+,+,+,+,+)$& $m'$ & $(+,-,-,-,-,-,-)$\\
$-h_i$ & $(+,{\stackrel{\mathsmaller{i}}{\overbrace{-,+,+}}},+,+,+)$ & $-\bar{f}'_i$ &
$(-,{\stackrel{\mathsmaller{i}}{\overbrace{+,-,-}}},-,-,-)$\\
$e_j$ & $(+,+,+,+,{\stackrel{\mathsmaller{j}}{\overbrace{-,+,+}}})$ & $\bar{a}'_j$ &
$(-,-,-,-,{\stackrel{\mathsmaller{j}}{\overbrace{+,-,-}}})$\\
$\bar{h}'_i$ &
$(+,{\stackrel{\mathsmaller{i}}{\overbrace{+,-,-}}},-,-,-)$&$f_i$ & 
$(-,{\stackrel{\mathsmaller{i}}{\overbrace{-,+,+}}},+,+,+)$\\
$q'_j$ &
$(+,-,-,-,{\stackrel{\mathsmaller{j}}{\overbrace{+,-,-}}})$&$a_j$ & 
$(-,+,+,+,{\stackrel{\mathsmaller{j}}{\overbrace{-,+,+}}})$\\
$\bar{g}'_{ji}$ &
$(-,{\stackrel{\mathsmaller{i}}{\overbrace{+,-,-}}},{\stackrel{\mathsmaller{j}}{\overbrace{+,-,-}}})$&
$b_{ji}$ & $(+,{\stackrel{\mathsmaller{i}}{\overbrace{-,+,+}}},{\stackrel{\mathsmaller{j}}{\overbrace{-,+,+}}})$\\
$a'_j$ &
$(-,-,-,-,{\stackrel{\mathsmaller{j}}{\overbrace{-,+,+}}})$&$q_j$ & 
$(+,+,+,+,{\stackrel{\mathsmaller{j}}{\overbrace{+,-,-}}})$\\
$-g_{ji}$ & 
$(-,{\stackrel{\mathsmaller{i}}{\overbrace{-,+,+}}},{\stackrel{\mathsmaller{j}}{\overbrace{-,+,+}}})$&
$-\bar{b}'_{ji}$ &
$(+,{\stackrel{\mathsmaller{i}}{\overbrace{+,-,-}}},{\stackrel{\mathsmaller{j}}{\overbrace{+,-,-}}})$\\
$-\bar{a}_j$&$(-,+,+,+,{\stackrel{\mathsmaller{j}}{\overbrace{+,-,-}}})$&
$-e'_j$&$(+,-,-,-,{\stackrel{\mathsmaller{j}}{\overbrace{-,+,+}}})$\\
$-\bar{b}_{ji}$&
$(+,{\stackrel{\mathsmaller{i}}{\overbrace{-,+,+}}},{\stackrel{\mathsmaller{j}}{\overbrace{+,-,-}}})$&
$-g'_{ji}$&$(-,{\stackrel{\mathsmaller{i}}{\overbrace{+,-,-}}},{\stackrel{\mathsmaller{j}}{\overbrace{-,+,+}}})$\\
$-m$&$(+,+,+,+,-,-,-)$&$-h'_0$&$(-,-,-,-,+,+,+)$\\
$b'_{ji}$&
$(+,{\stackrel{\mathsmaller{i}}{\overbrace{+,-,-}}},{\stackrel{\mathsmaller{j}}{\overbrace{-,+,+}}})$&
$\bar{g}_{ji}$&
$(-,{\stackrel{\mathsmaller{i}}{\overbrace{-,+,+}}},{\stackrel{\mathsmaller{j}}{\overbrace{+,-,-}}})$\\
$f'_i$&
$(-,{\stackrel{\mathsmaller{i}}{\overbrace{+,-,-}}},+,+,+)$&
$\bar{h}_i$&$(+,{\stackrel{\mathsmaller{i}}{\overbrace{-,+,+}}},-,-,-)$\\
$-e'_0$&$(+,-,-,-,+,+,+)$&$-\bar{h}_0$&$(-,+,+,+,-,-,-)$\\
$-\bar{f}_i$&$(-,{\stackrel{\mathsmaller{i}}{\overbrace{-,+,+}}},-,-,-)$&
$-h'_i$&$(+,{\stackrel{\mathsmaller{i}}{\overbrace{+,-,-}}},+,+,+)$
\end{tabular}
\end{center}
\caption{Spinorial embedding of the background fluxes. The weights in each column correspond to 
one of the two Weyl spinors on which the set of fluxes ${\mathbb G}$ can be decomposed.} \label{mapa}
\end{table}

One can proceed analogously with the set of moduli ${\mathbb T}$. In 
this case they transform as a vectorial ${\mathbf 7}$ of $SL(2,\Z)^7$, 
as shown in table \ref{pesosmoduli}. Let us define
\beq
e^{i\mathbb T}~\equiv~ 1+ i{\mathbb T}-{\mathbb T}\otimes{\mathbb T}+\ldots  \ .
\eeq
In this language, the superpotential (\ref{todo}) then takes 
the very compact form
\beq
W={\mathbb G}\otimes e^{i\mathbb T}\vert_{(+,+,+,+,+,+,+)}  \ ,
\label{wG}
\eeq
which is reminiscent of the typical expressions for flux induced superpotentials.

\begin{table}[!ht]
\begin{center}
\begin{tabular}{c|c}
Moduli & Weight\\
\hline \hline $S$ & $(1,0,0,0,0,0,0)$\\
$T_i$ & $(0,{\stackrel{\mathsmaller{i}}{\overbrace{1,0,0}}},0,0,0)$\\
$U_i$ & $(0,0,0,0,{\stackrel{\mathsmaller{i}}{\overbrace{1,0,0}}})$
\end{tabular}
\end{center}
\caption{Embedding of the moduli in a $\mathbf 7$ of $SL(2,\Z)^7$.} \label{pesosmoduli}
\end{table}

Moreover, the Bianchi identities now correspond to constraints in the components of the
bispinor of fluxes
\beq
{\mathbb G}\otimes{\mathbb G}~=~{\mathbb G}\cdot{\mathbb G}~\oplus~ 
{\mathbb G}\Gamma^{I_1}{\mathbb G}~\oplus~\ldots~\oplus~{\mathbb G}
\Gamma^{I_1I_2I_3I_4I_5I_6I_7}{\mathbb G}  \  ,
\label{GtG}
\eeq
where $\Gamma^{I_1\ldots I_n}\equiv \Gamma^{[I_1}\cdot\ldots\cdot\Gamma^{I_n]}$ 
and $\Gamma^I$ are the complexified gamma matrices of the relevant Clifford 
algebra, and $I_a=1,\bar{1},\ldots,7,\bar{7}$. 

\end{document}